\documentclass[aps,pre,
   11pt,
   preprint,
   tightenlines,
   superscriptaddress,
   eqsecnum,
   floatfix,
   showpacs,
   preprintnumbers,
   longbibliography,
   nofootinbib
   ]{revtex4-2}
\usepackage{dcolumn}    
\usepackage{bm}         
\usepackage{color}
\usepackage{framed}
\usepackage{subfigure}
\usepackage{stackrel}
\usepackage{amsmath}
\usepackage{amssymb}
\usepackage{time}
\usepackage{graphicx}
\usepackage{tikz}
\usetikzlibrary{calc,trees,positioning,arrows,chains,shapes.geometric,%
    decorations.pathreplacing,decorations.pathmorphing,shapes,%
    matrix,shapes.symbols}
\tikzset{
>=stealth',
  punktchain/.style={
    rectangle, 
    rounded corners, 
    draw=black, very thick,
    text width=10em, 
    minimum height=2em, 
    text centered, 
    on chain},
  line/.style={draw, thick, <-},
  element/.style={
    tape,
    top color=white,
    bottom color=blue!50!black!60!,
    minimum width=8em,
    draw=blue!40!black!90, very thick,
    text width=10em, 
    minimum height=3.5em, 
    text centered, 
    on chain},
  every join/.style={->, thick,shorten >=1pt},
  decoration={brace},
  tuborg/.style={decorate},
  tubnode/.style={midway, right=2pt},
}
\DeclareSymbolFont{AMSb}{U}{msb}{m}{n}
\DeclareSymbolFontAlphabet{\mathbb}{AMSb}
\usepackage{physics}
\newcommand{\ef}[1]{\eqref{#1}}                    
\newcommand{\Schrodinger}{Schr{\"o}dinger} 
 
\newcommand{\etal}{\textit{et~al.}}        
\newcommand{\ansatz}{ans{\"a}tz}           
\newcommand{\rmi}{{\rm i}}                     
\newcommand{\rme}{{\rm e}}                     
\newcommand{\DD}[1]{\mathcal{D}{#1} \, }       
\newcommand{\bzero}{\vb{0}}
\newcommand{\expB}[1]{\exponential \Bigl \{ \, #1 \, \Bigr \}}
\newcommand{\calR}{\mathcal{R}}
\newcommand{\calI}{\mathcal{I}}
\usepackage{hyperref}
\hypersetup{
    pdfnewwindow=true,      
    colorlinks=true,        
    linkcolor=red,          
    citecolor=blue,         
    filecolor=green,        
    urlcolor=cyan           
}
\begin{document}
\preprint{}
%
%
\title{Modeling pandemics}
\author{John F. Dawson}
\email[Email: ]{john.dawson@unh.edu}
\affiliation{Department of Physics,
   University of New Hampshire,
   Durham, NH 03824, USA}
\author{Fred Cooper} 
\email[Email: ]{cooper@santafe.edu}
\affiliation{The Santa Fe Institute, 
   1399 Hyde Park Road, 
   Santa Fe, NM 87501, USA}
\affiliation{Theoretical Division,
   Los Alamos National Laboratory,
   Los Alamos, NM 87545, USA}
\author{Efstathios G. Charalampidis}  
\email[Email: ]{echaralampidis@sdsu.edu}
\affiliation{Nonlinear Dynamical Systems Group,
   Computational Sciences Research Center, and
   Department of Mathematics and Statistics,
   San Diego State University,
   San Diego, CA 92182-7720, USA}
\date{\today, \now \ PST}
%
%
\begin{abstract}
We review several models for pandemics that plagued the USA and the world in the past decade.  Methods of data fitting are reviewed and several types of microscopic and rate equation models are discussed and numerically solved.  This paper was written in March of 2021 and newer data is now available; however some of the models and techniques we used to numerically study these models are still of interest.  Several appendices discuss in detail these models.  
\begin{figure}[h]
\centering
\includegraphics[width=0.55\textwidth]{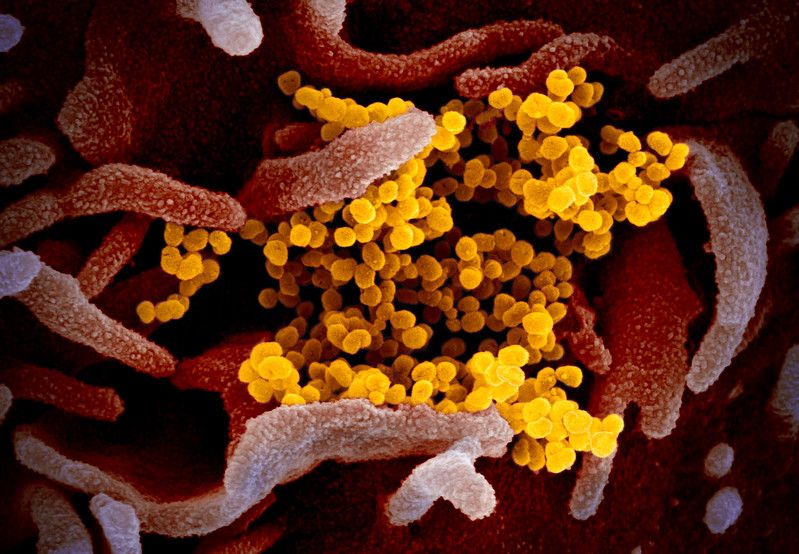}
\end{figure}

Scanning electron microscope image of the SARS-CoV-2 (COVID-19) virus (yellow) among human cells (pink) taken at the Rocky Mountain Laboratories\footnote{Released by the Rocky Mountain Laboratories (RML) at the National Institute of Allergy and Infectious Diseases (NIH), Hamilton, Montana.}.  The image was isolated from a patient in the US.  
\end{abstract}
%
%
%
\maketitle   
\newpage
\tableofcontents
%
%
\newpage
\section{\label{s:intro}Introduction}

As a result of the COVID-19 pandemic, numerous mathematical models of the pandemic have recently been proposed and debated in both the press and in scientific publications.  One way to think about these models are to categorize them according to the way they work.  One can divide them into two major groups: those models that simply try to predict future conditions based on statistical models fit to past conditions, and those that are based on transition dynamics.  In the first category is William Farr's original proposal (1844) that epidemics follow a bell-shaped curve so that after peak infection is reached, the epidemic will ``progressively fall through the same steps.''  One fits the data to a normal distribution curve.  Obviously one can choose a different model curve for the epidemic and simply fit your favorite curve to the data or to the log of the data.  A recent model by Morte \etal\ \cite{10.3389/fphy.2020.00144} chooses a model curve, inspired by renormalization group techniques, and fits the curve to the observed number of cases available to date.  The model from the Institute for Health Metrics and Evaluation (IHME) at UW appears to belong to this first category \cite{10.7326/M20-1565}, although apparently modified somewhat by social distancing models.  The model used mortality data rather than infection rate data which is considered by some to be more reliable.  The IHME model has been widely quoted in the press and by the president's task force, largely because they are willing to make predictions.  

In the second category are epidemiological models often based on SEIR (susceptible-exposed-infectious-recovered) dynamics, with numerous modifications to account for social distancing.  We discuss these types of models in Sec.~\ref{ss:SEIR}.  A recent model by Dandekar and Barbastathis \cite{Dandekar2020.04.03.20052084} (the ``MIT'' model) using machine learning techniques belongs to this category.  We discuss this model below.  The MRC center of the Imperial College of London model is also based on epidemiological considerations as well as statistical analysis.  

A third method (Girona \cite{10.3389/fphy.2020.00186}) applies stochastic algorithms to individuals players in a sample population to create the pandemic in a two-dimensional grid.  This model is based on reaction rates and is discussed below.

%
%
\section{\label{s:data}Current US COVID-19 data}

\begin{figure}[h]
  \centering
  \subfigure[\ US cases]
  { \label{f:UScases}
    \includegraphics[width=0.45\columnwidth]{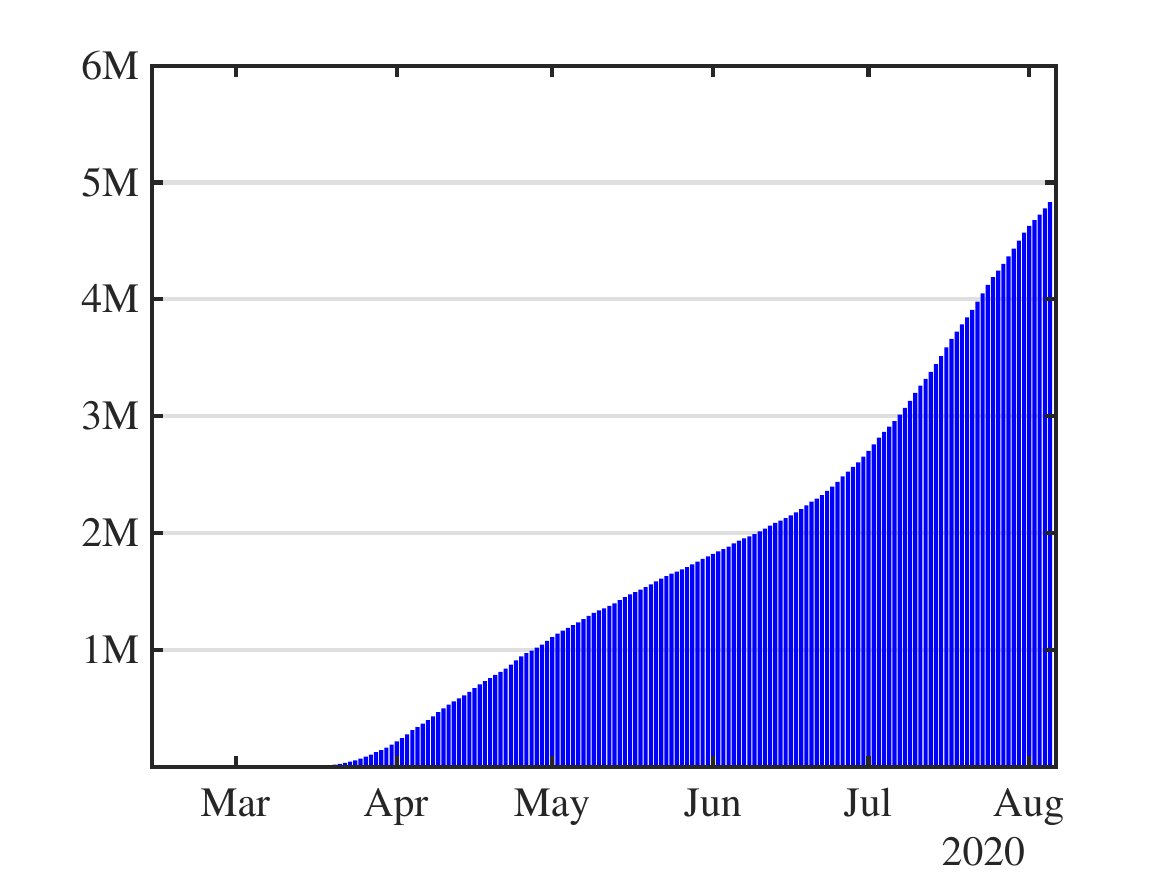} }
  \subfigure[\ US deaths]
  { \label{f:USdeaths}
    \includegraphics[width=0.45\columnwidth]{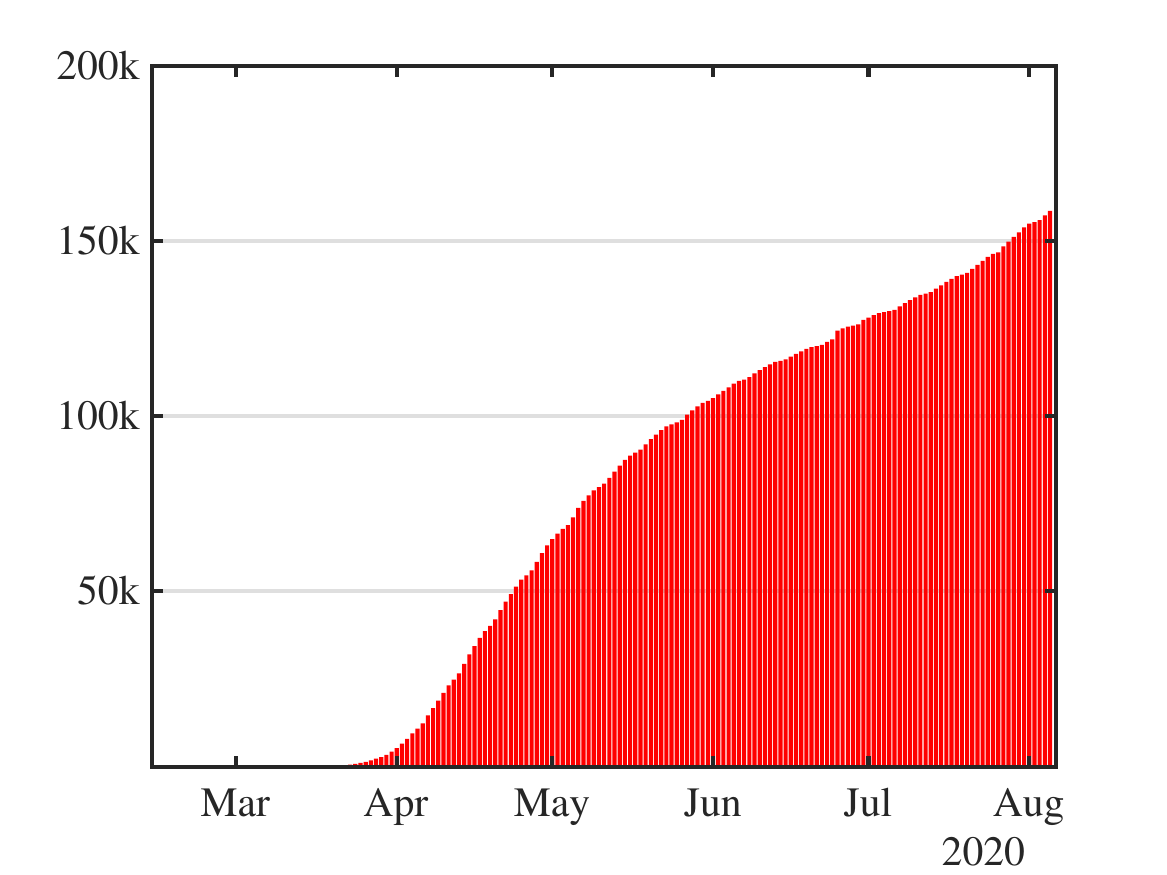} }
  \subfigure[\ new US cases]
  { \label{f:newUScases}
     \includegraphics[width=0.45\columnwidth]{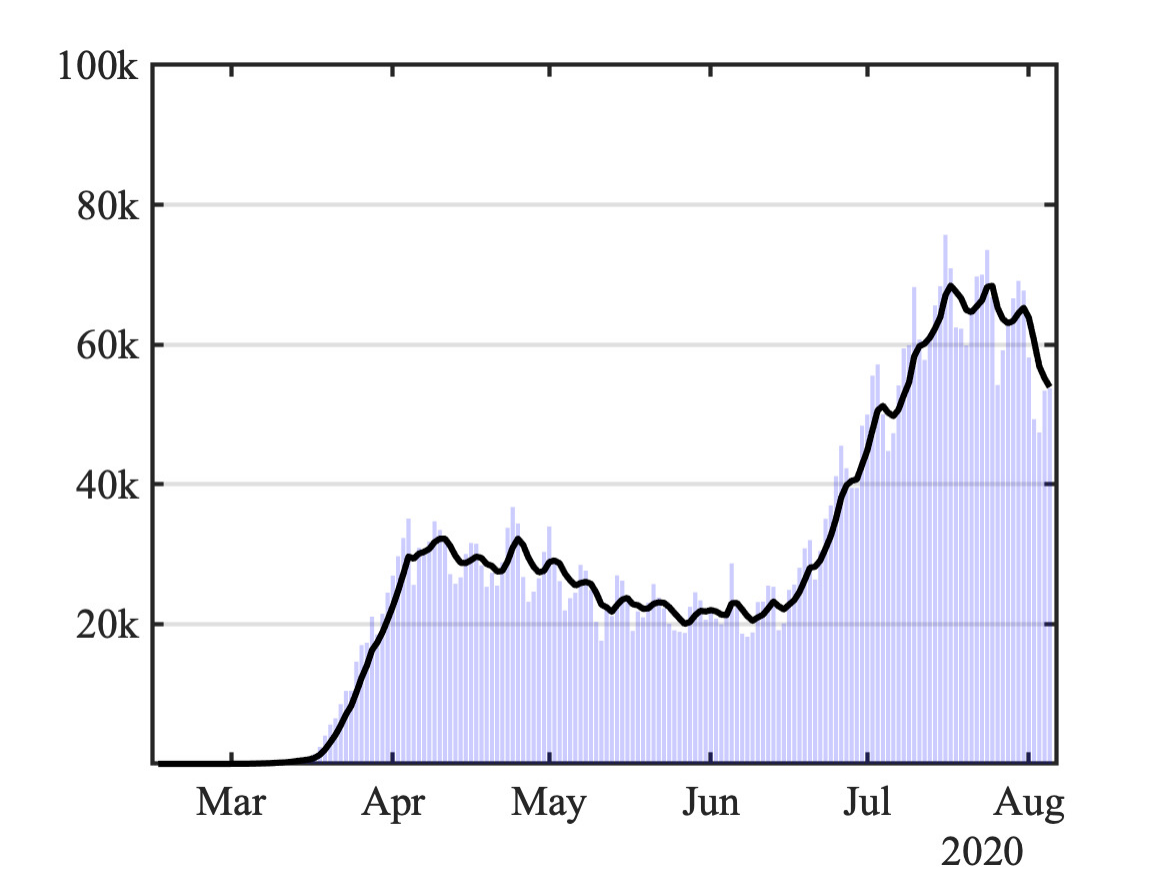} }
  \subfigure[\ new US deaths]
  { \label{f:newUSdeaths}
     \includegraphics[width=0.45\columnwidth]{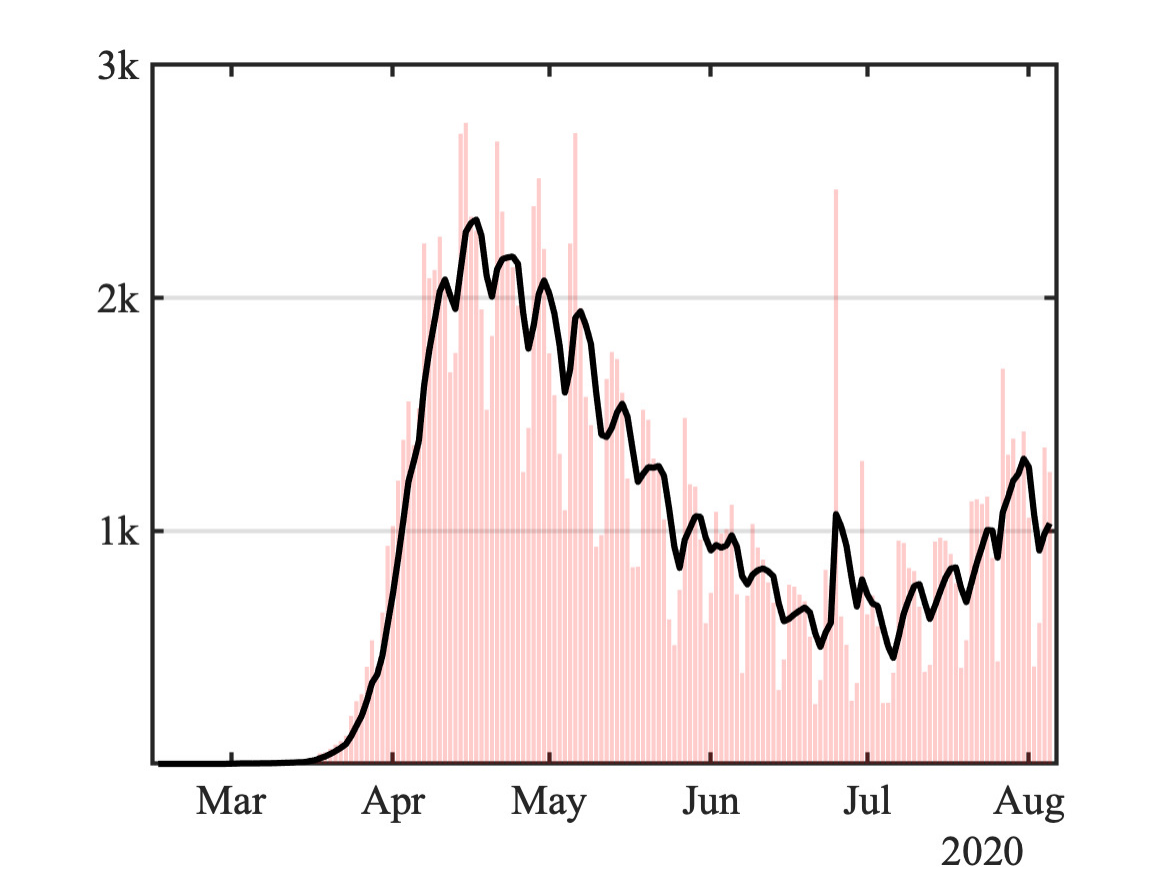} }
  \caption{\label{f:NYT-USdata}Data from the NYT database for COVID-19 
  cases in the US as of June 25, 2020.  The black lines are seven day 
  moving averages.}
\end{figure}

A plots of confirmed cases, deaths, new cases, and new deaths from COVID-19 in the US as of June 25, 2020 is shown in Fig.~\ref{f:NYT-USdata}.  Data is from the New York Times data base: \url{https://github.com/nytimes/covid-19-data}.  The population of the United States is about 330M and the density is about 100 people per mi${}^{2}$ overall.  The black lines in Figs.~\ref{f:newUScases} and \ref{f:newUSdeaths} are seven day moving averages.  Most experts believe that the real number of cases and deaths are higher than the reported numbers.  Note the upswing in new cases on about June 15th.

%
%
%
%

%
%
\section{\label{s:models}Models}

We discuss various models in this section. 

%
%
\subsection{\label{ss:datafitting}Data fitting models}

%
%
\begin{figure}[ht]
  \centering
  \subfigure[\ log of US cases]
  { \label{f:LogUScases}
    \includegraphics[width=0.45\columnwidth]{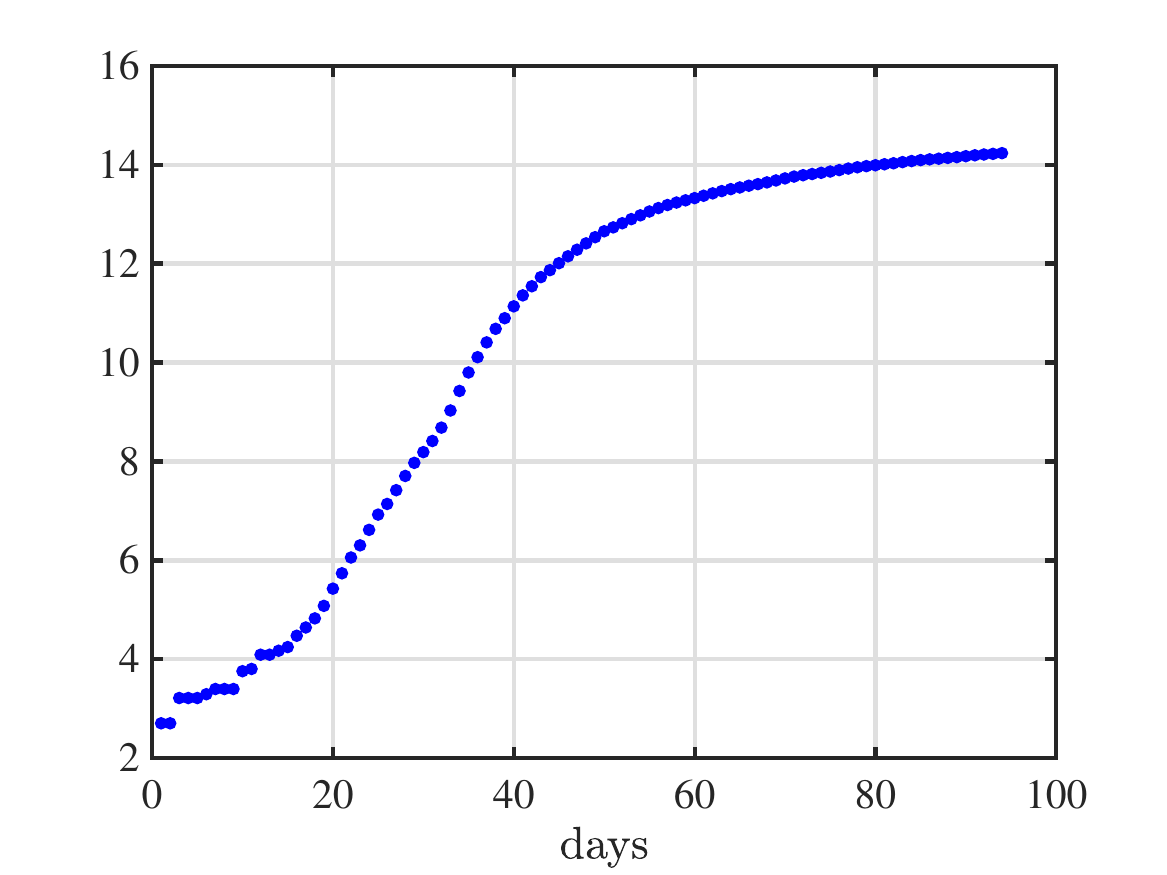} }
  \subfigure[\ US cases]
  { \label{f:RenormUScases}
    \includegraphics[width=0.45\columnwidth]{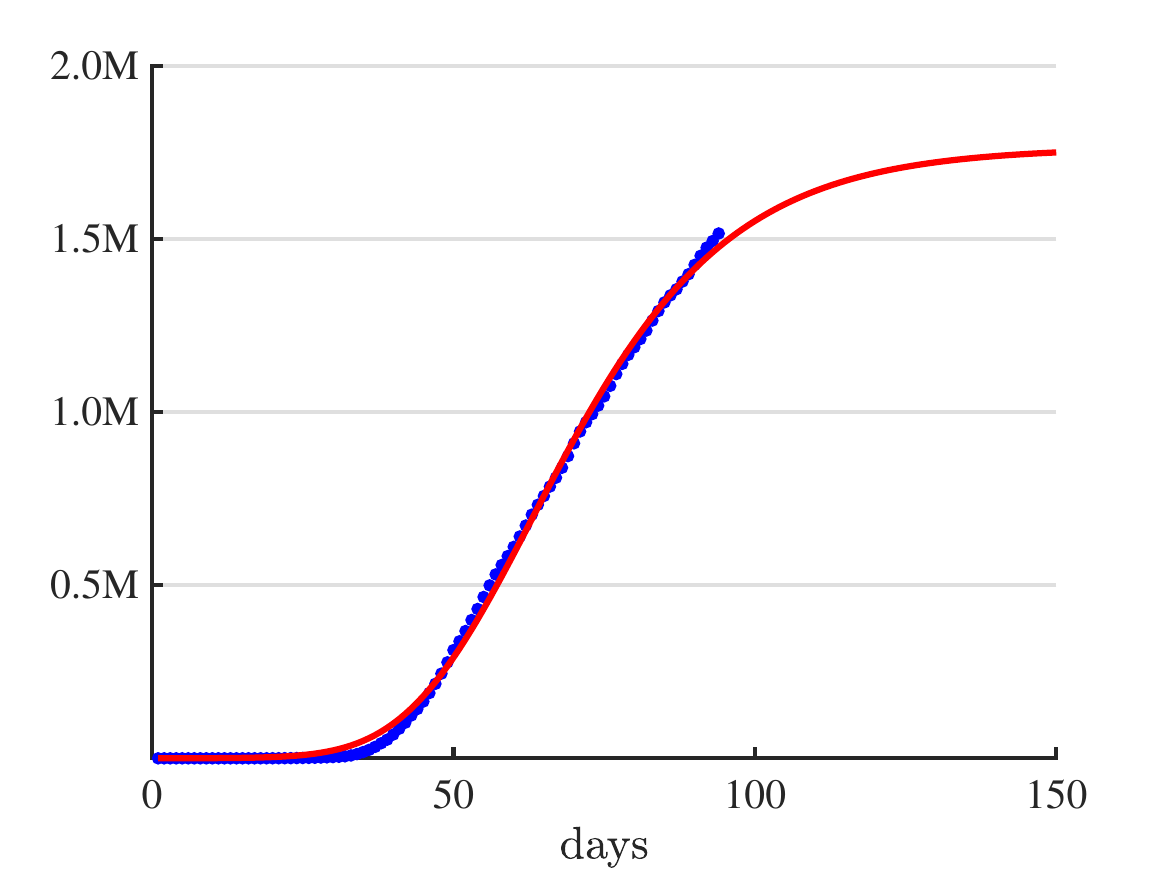} }
  \caption{\label{f:Renorm} (a) Plot of the log of the number of
  US cases and (b) fit of the renormalization \ansatz\ to US cases as
  of May 18, 2020.}
\end{figure}
\begin{figure}[ht]
  \centering
  \includegraphics[width=0.75\columnwidth]{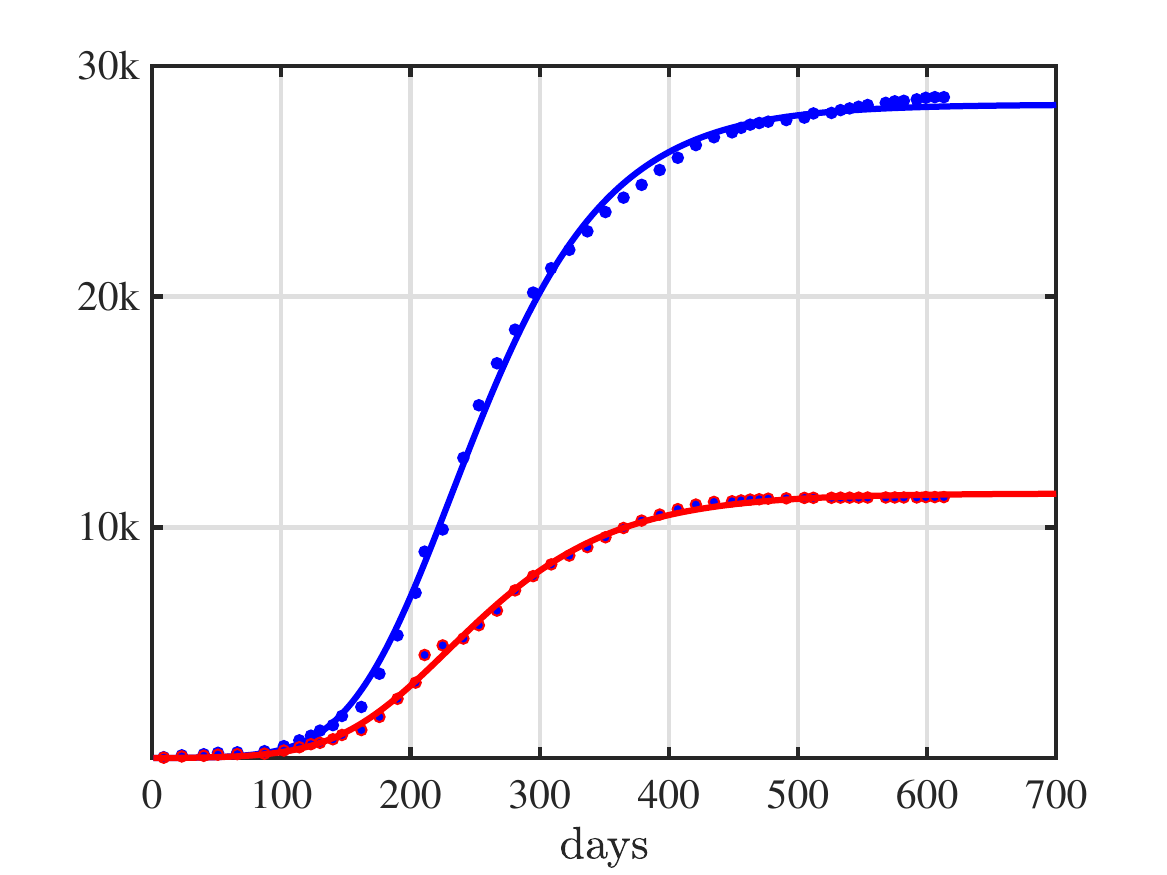}
  \caption{\label{f:EbolaFitRenorm} Fit of the exponential Fermi-Dirac distribution
  \ef{e:FermiDirac} to the number of cases and number of deaths in the 
  Ebola epidemic in West Africa in 2014-2915.}
\end{figure}

%
%
We briefly describe the fitting method of Morte \etal\ \cite{10.3389/fphy.2020.00144}.  These authors noted that the log of the number of cases, shown in Fig.~\ref{f:LogUScases} appears to look like a Fermi-Dirac distribution, given by:
\begin{equation}\label{e:FermiDirac}
   \alpha(t)
   =
   \frac{a}{1 + \exp{-\gamma (t - t_0) }} \>.
\end{equation}
So one might be able to fit the US cases data by an \ansatz\ of the form: $I(t) = \exp{ \alpha(t) }$, by adjusting the three parameters: $(a, t_0, \gamma)$.  This in fact seems to be the case, as shown in Fig.~\ref{f:RenormUScases} where the fit was done using the {\tt lsqcurvefit} routine in MatLab.  The fit was much better when fit to the number of cases rather than the log of the number of cases, and looks very good with the parameters $(a, t_0, \gamma) = (14.309,16.887,0.0599)$.  Renormalization ideas are only used to relate the \ansatz\ to a rate equation, not for the actual fitting of the data.  The rate equation involves powers of $I(t)$, and does not really give much insight about the underlying dynamics.  So this method belongs to the first category of models.  Whereas fits to current data in Fig.~\ref{f:RenormUScases} looks good, one wonders if the US data has really reached a point where predictions of this type can be relied upon.  However based on the current data, it predicts about 1.75 million total infections in the US, however this number keeps changing when on more data becomes available.  In other countries which have progressed further along, it seems to work much better.  An example is the fit to the Ebola epidemic in West Africa in 2014-2015, shown in Fig.~\ref{f:EbolaFitRenorm}, which seems to fit the data very well, including the total number of cases and deaths during the epidemic.

%
%
%
%

%
%
\subsection{\label{ss:SEIR}SEIR models}

\begin{figure}[ht]
  \centering
  \subfigure[\ Linearized solutions]
  { \label{f:SEIRlinearAll}
    \includegraphics[width=0.45\columnwidth]{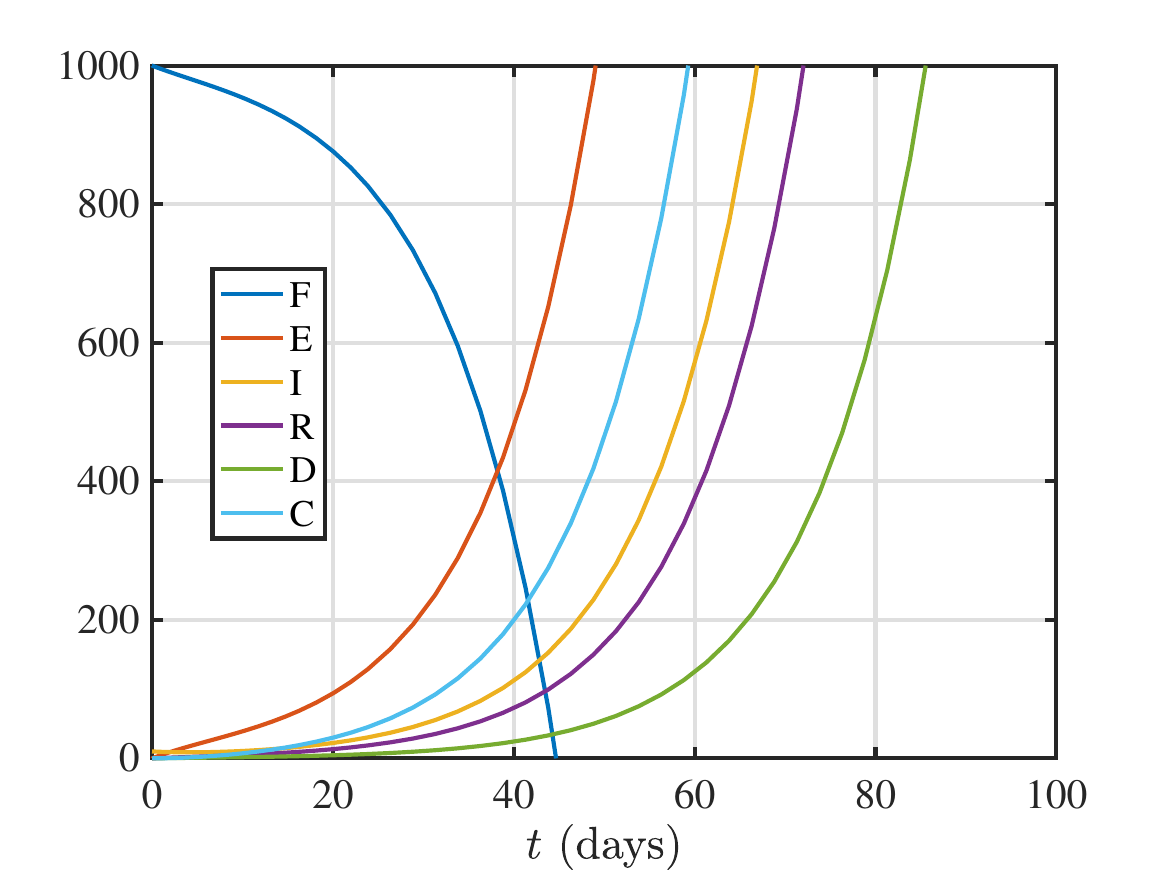} }
  \subfigure[\ New cases and deaths]
  { \label{f:SEIRlinearNew}
    \includegraphics[width=0.45\columnwidth]{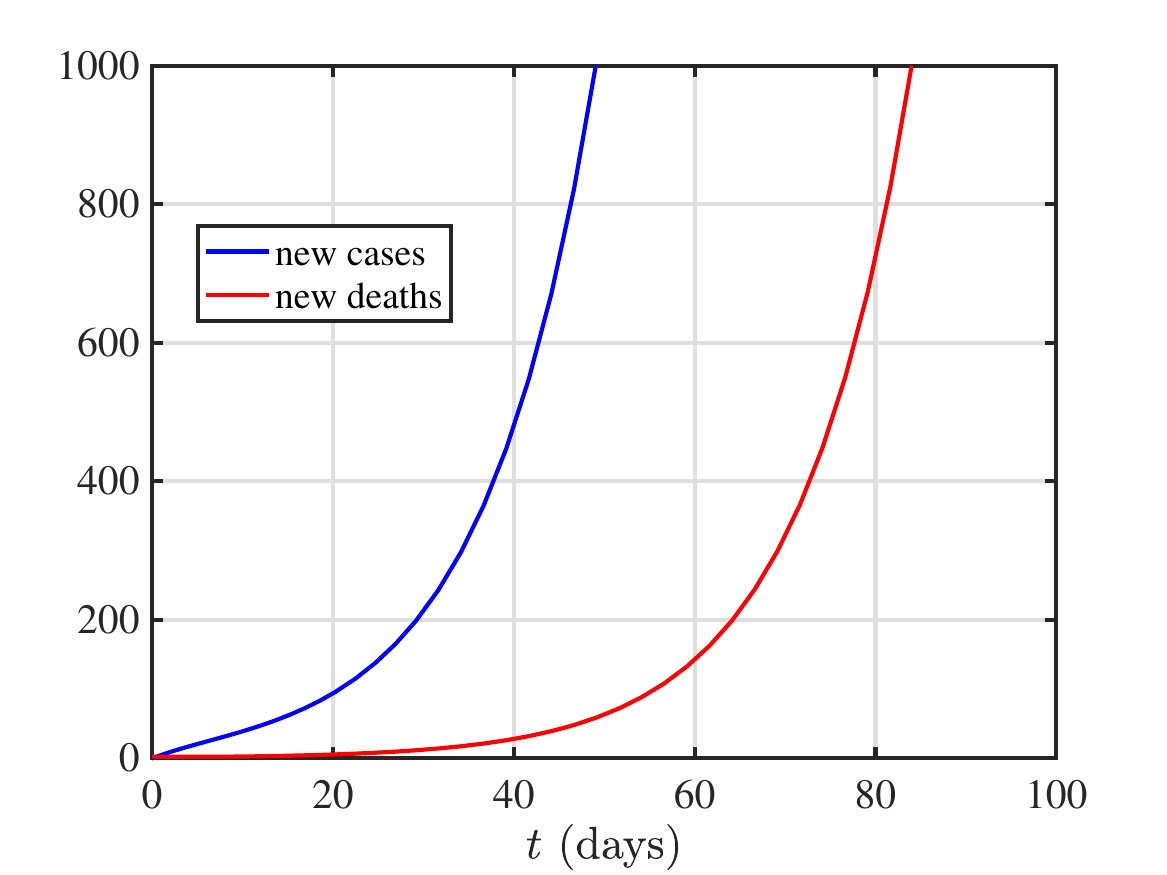} }
  \caption{\label{f:SEIRlinear}Plots of solutions of the linearized equations
  \ef{e:SEIRII} for the case when $\beta = 1/2$, $\sigma = 1/24$, and 
  $\gamma = 1/14$, with the initial values $F_0 = 1000$, $E_0 = 0$, 
  $I_0 = 10$, $R_0 = D_0 = C_0 = 0$, and $f = 0.25$.}
  
\end{figure}

The basic SEIR model assumes a homogeneous population described by 
the following consituants:
\begin{center}
\begin{tabular}{c@{\qquad}l}
\hline
\emph{Symbol} & \emph{Population}\\
$S$ & Susceptible individuals \\
$E$ & Exposed individuals \\
$I$ & Infected individuals \\
$R$ & Removed individuals \\
\hline
\end{tabular}
\end{center}
with reactions and reaction rates given by:
\begin{equation}\label{e:reactions}
   S + I 
   \xrightarrow{\beta} 
   E + I
   \qc
   E
   \xrightarrow{\sigma}
   I
   \qc
   I
   \xrightarrow{\gamma} 
   R \>.
\end{equation}
Here after contact with an infected individual ($I$), susceptible ($S$) individuals enter into the exposed class ($E$) at a transmission rate $\beta$ per person which progress to the infected class ($I$) at an infection or incubation rate $\sigma$ and which in turn are eventually removed ($R$) from the dynamics at a rate $\gamma$.  In this model, the removed either become immune or die.  Rate equations for these processes are:
\begin{subequations}\label{e:SEIRbasic}
\begin{align}
   \dv{S}{t}
   &=
   - \beta \, S(t) I(t) / N \>,
   \label{e:SEIRbasic-a} \\
   \dv{E}{t}
   &=
   \beta \, S(t) I(t) / N - \sigma \, E(t) \>,
   \label{e:SEIRbasic-b} \\
   \dv{I}{t}
   &=
   \sigma \, E(t) - \gamma \, I(t) \>,
   \label{e:SEIRbasic-c} \\
   \dv{R}{t}
   &=
   \gamma \, I(t) \>,
   \label{e:SEIRbasic-d}
\end{align}
\end{subequations}
where the total population number $N = S(t) + E(t) + I(t) + R(t)$ is conserved.  
There are three parameters in this model, $(\beta,\sigma,\gamma)$, in addition to the total population number $N$.  One can split the removed category into a recovered (now called $R$) and dead ($D$) by replacing Eq.~\ef{e:SEIRbasic-d} with the equations,
\begin{equation}\label{e:RDdynamics}
   \dv{D}{t}
   =
   f \gamma \, I(t)
   \qc
   \dv{R}{t}
   =
   (1 - f) \, \gamma \, I(t) \>,
\end{equation}
where $f$ is the fraction of infected individuals that die. The recovered are assumed to be immune.  The total number of individuals are now given by $N = S(t) + E(t) + I(t) + R(t) + D(t)$, and include the dead.  The cumulative number of infected individuals $C(t)$ are given by $C(t) = I(t) + D(t) + R(t)$, and the new infections per day by
\begin{equation}\label{e:dCdt}
   \dv{C}{t}
   = 
   \sigma \, E(t) \>.
\end{equation}
For the COVID-19 case, the transmission time $1/\beta$ is approximately 2--3 days if no control measures are taken.  The incubation time $1/\sigma$ is approximately 7--14 days and the recovery or death time $1/\gamma$ also seems to be about 7--14 days.  The death rate in the US is about 5\%.  For the Ebola epidemic in West Africa, the death rate was a shocking 50\%.  The basic infection (reproduction) number is defined to be $\calR_0 = \beta/\gamma$ and the basic incubation number is $\calI_0 = \sigma/\gamma$.  The initial conditions are also important.  Here $N$ is the total population number which in the case of the US is about 330 million people.  For New York City it is about 8 million, for Wuhon China about 11 million.  These are big numbers compared to the number of infected individuals at $t=0$, which can be as small as 50 infected persons.  Choosing $S_0 = N - I_0$, we see that until a significant fraction of the initial population becomes infected, $S(t) \approx S_0 \approx N$.  So it is a reasonable to set $S(t) = N + F(t)$, where $F(t)$ is a number comparable to the populations of exposed, infected, and removed individuals.  With this approximation, the SEIR equations \ef{e:SEIRbasic} become:
\begin{subequations}\label{e:SEIRII}
\begin{align}
   \dv{F}{t}
   &=
   - \beta \, I(t) \>,
   \label{e:SEIRII-a} \\
   \dv{E}{t}
   &=
   \beta \, I(t) - \sigma \, E(t) \>,
   \label{e:SEIRII-b} \\
   \dv{I}{t}
   &=
   \sigma \, E(t) - \gamma \, I(t) \>,
   \label{e:SEIRII-c} \\
   \dv{D}{t}
   &=
   f \gamma \, I(t)
   \qc
   \dv{R}{t}
   =
   (1 - f) \, \gamma \, I(t) \>,
   \label{e:SEIRII-d}
\end{align}
\end{subequations}
Exact solutions of the linearized equations \ef{e:SEIRII} are given in Appendix~\ref{s:linearSEIR}.  Numerical solutions of the equations are illustrated in Fig.~\ref{f:SEIRlinear} for the case when $\beta = 1/2$, $\sigma = 1/24$, and $\gamma = 1/14$, with the initial values $F_0 = 1000$ (an arbitrary number to illustrate the trend), $E_0 = 0$, $I_0 = 10$, $R_0 = D_0 = C_0 = 0$, and $f = 0.25$.  These results apply to the initial onset of an epidemic, illustrating the exponential increase in cases and deaths.

%
%
%
%
%
%
\begin{figure}[ht]
  \centering
  \subfigure[\ Solutions of the SEIR equations]
  { \label{f:SEIRAllPlt}
    \includegraphics[width=0.45\columnwidth]{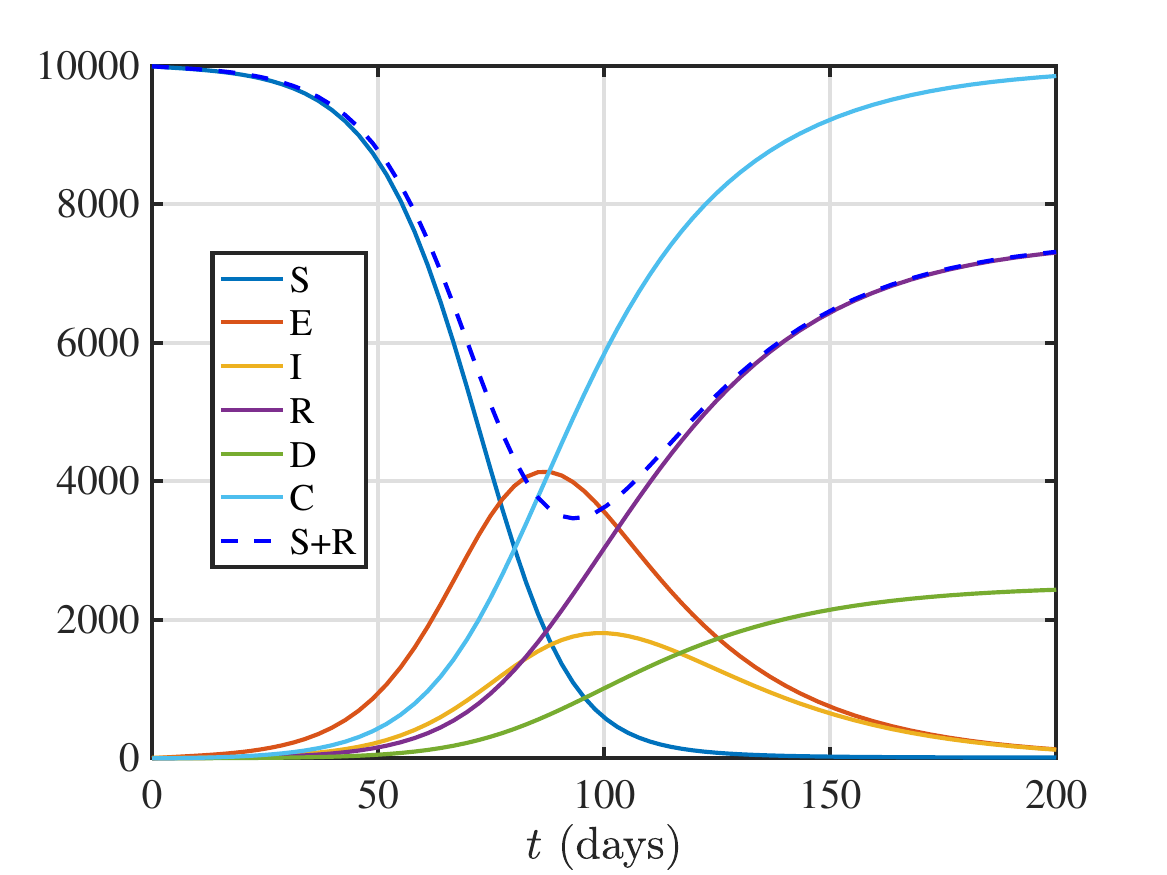} }
  \subfigure[\ SEIT new cases and deaths]
  { \label{f:SEIRNewCases}
    \includegraphics[width=0.45\columnwidth]{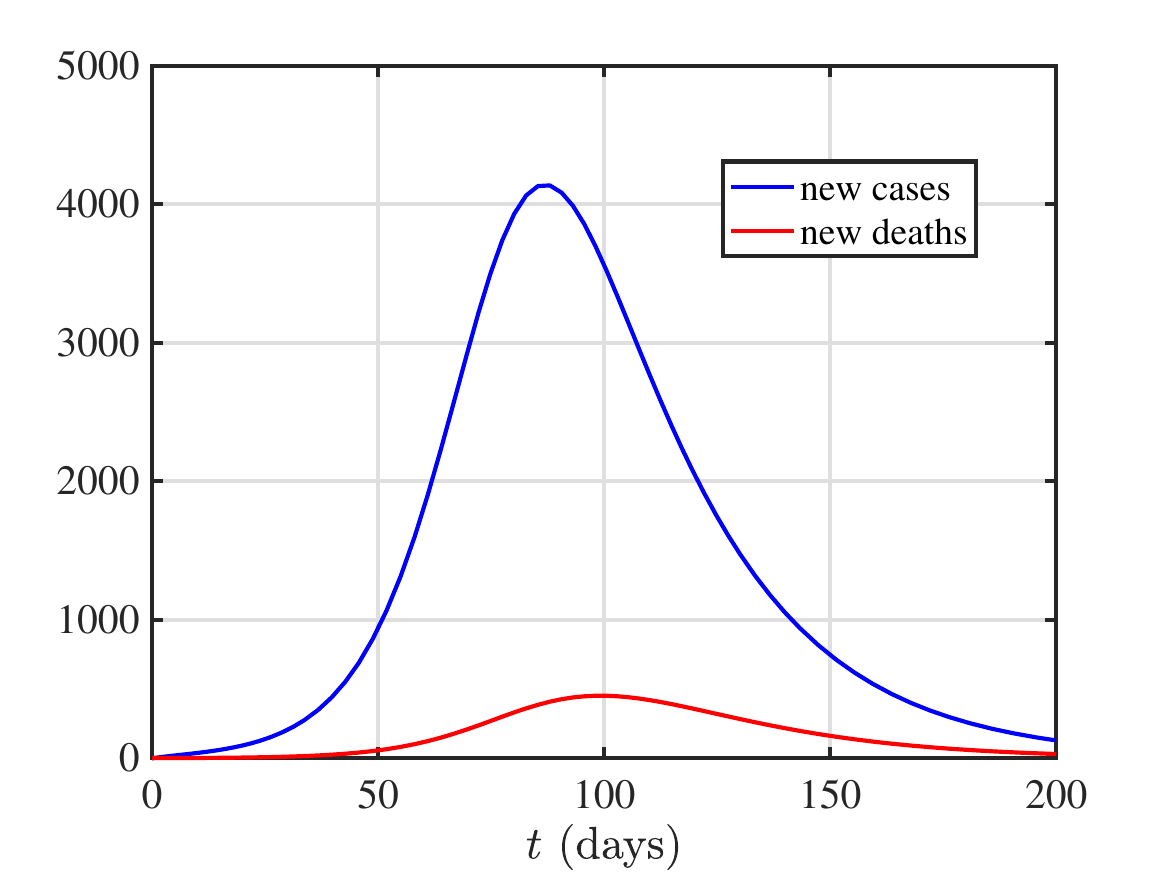} }
  \caption{\label{f:SEIRbasic}Plots of solutions of the SEIR equations
  \ef{e:SEIRbasic} for the case when $\beta = 1/2$, $\sigma = 1/24$, and 
  $\gamma = 1/14$, with the initial values $N_0 = 10000$, $S_0 = N_0 - I_0$,
  $E_0 = 0$, $I_0 = 10$, $R_0 = D_0 = C_0 = 0$, and $f = 0.25$.}
  
\end{figure}
%
%

In order to study the endpoint of an epidemic, we need to solve the complete SEIR model of \ef{e:SEIRbasic}.  This requires selecting a value for the total population number.  For these simulations, we selected a value of $N_0 = 10,000$, and used the same parameters as in Fig.~\ref{f:SEIRlinear}, with $\beta = 1/2$, $\sigma = 1/24$, and $\gamma = 1/14$, and with $S_0 = N_0 - I_0$, $E_0 = 0$, $I_0 = 10$, $R_0 = D_0 = C_0 = 0$, and $f = 0.25$.  The results are shown in Fig.~\ref{f:SEIRbasic}.
In this case approximately 5,000 individuals are infected and there are about 2,100 deaths out of an initial population of 10,000 individuals.  In the end, all the initial population is either infected, recovers, or dies.  This is due to the large reproduction rate here, $\calR_0 = \beta/\gamma = 7$.

\begin{figure}[ht]
  \centering
  \subfigure[\ SEIR solutions no controls]
  { \label{f:SEIRPlt-a}
    \includegraphics[width=0.45\columnwidth]{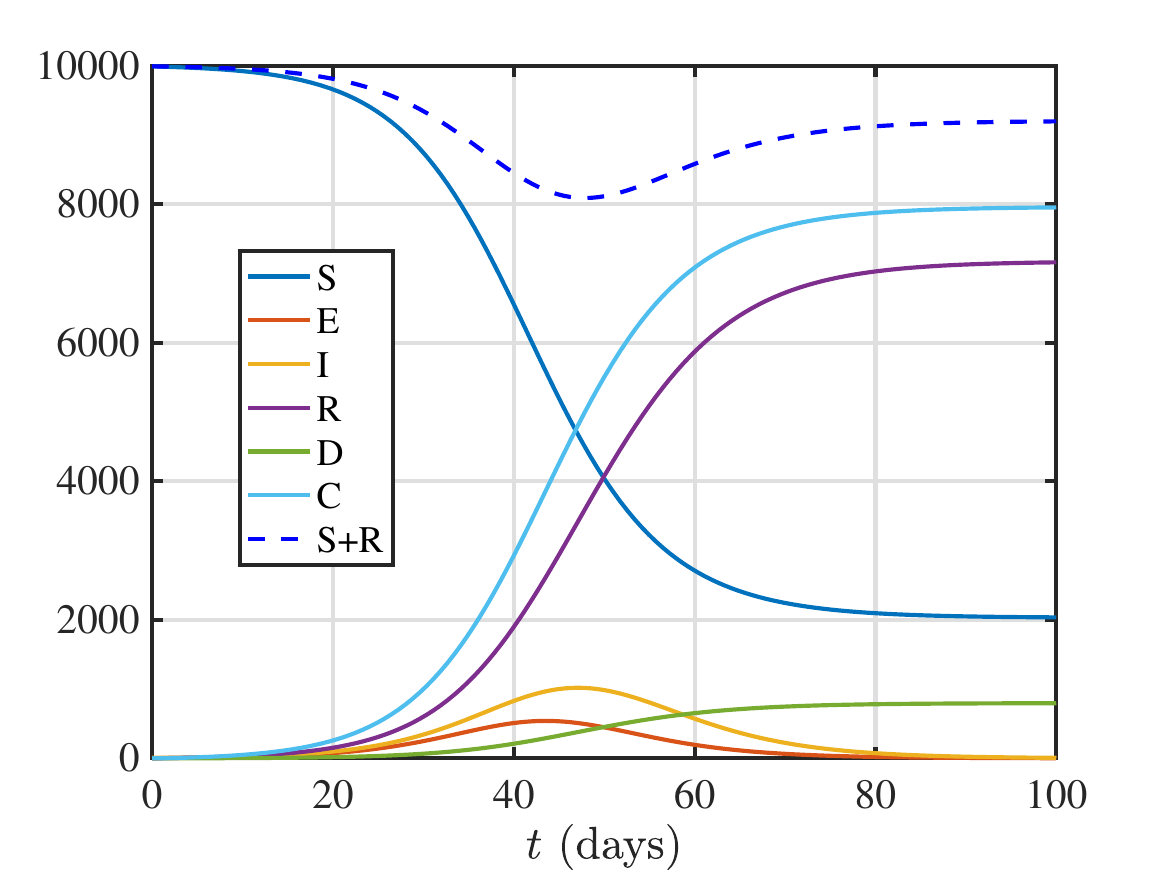} }
  \subfigure[\ SEIR new cases and deaths no controls]
  { \label{f:SEIRPlt-b}
    \includegraphics[width=0.45\columnwidth]{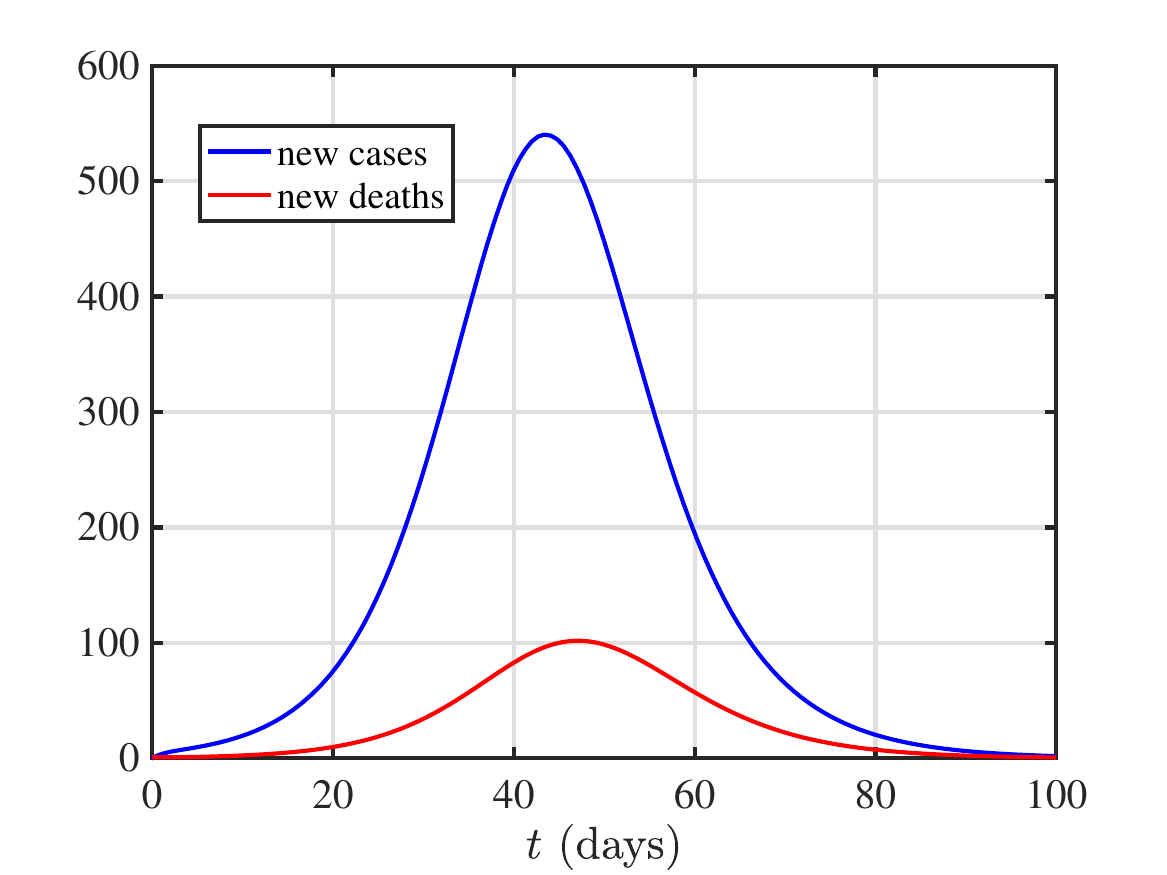} }
  \subfigure[\ SEIR solutions with controls]
  { \label{f:SEIRPlt-c}
    \includegraphics[width=0.45\columnwidth]{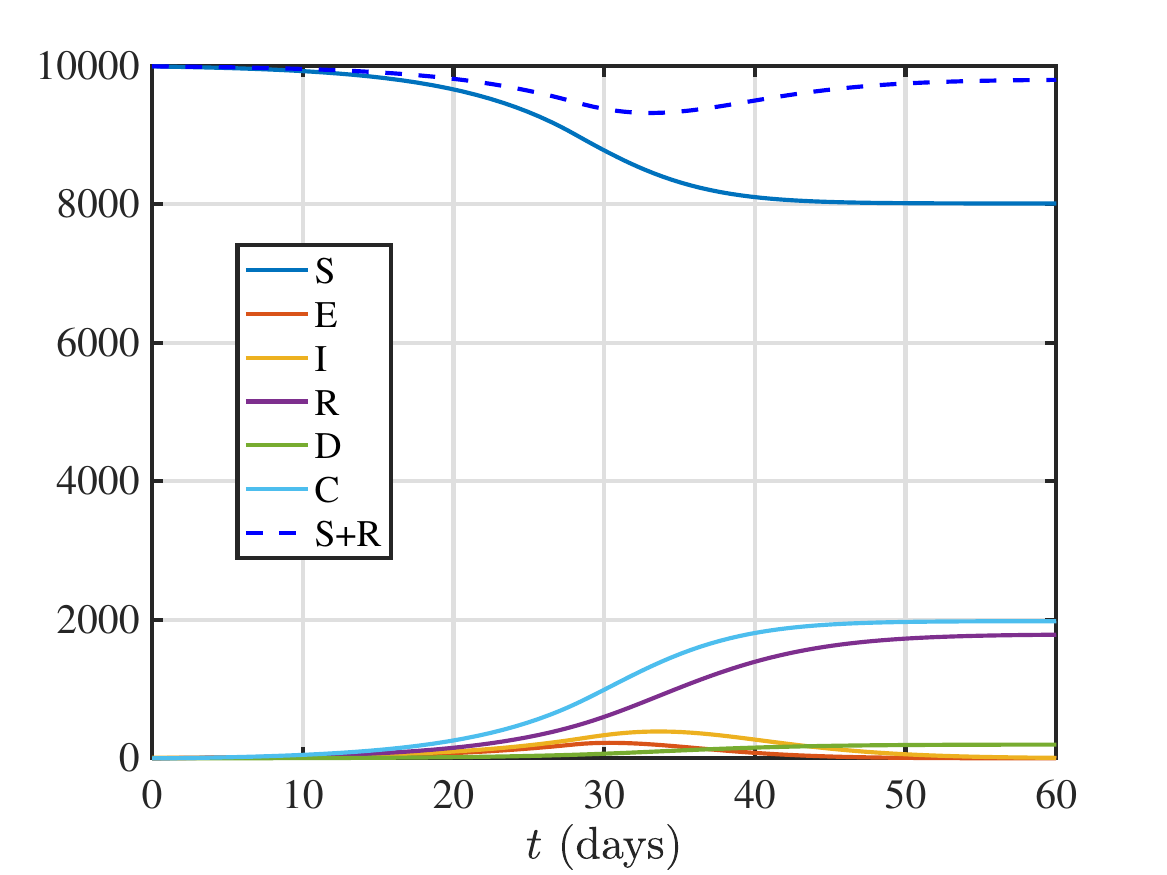} }
  \subfigure[\ SEIR new cases and deaths with controls]
  { \label{f:SEIRPlt-d}
    \includegraphics[width=0.45\columnwidth]{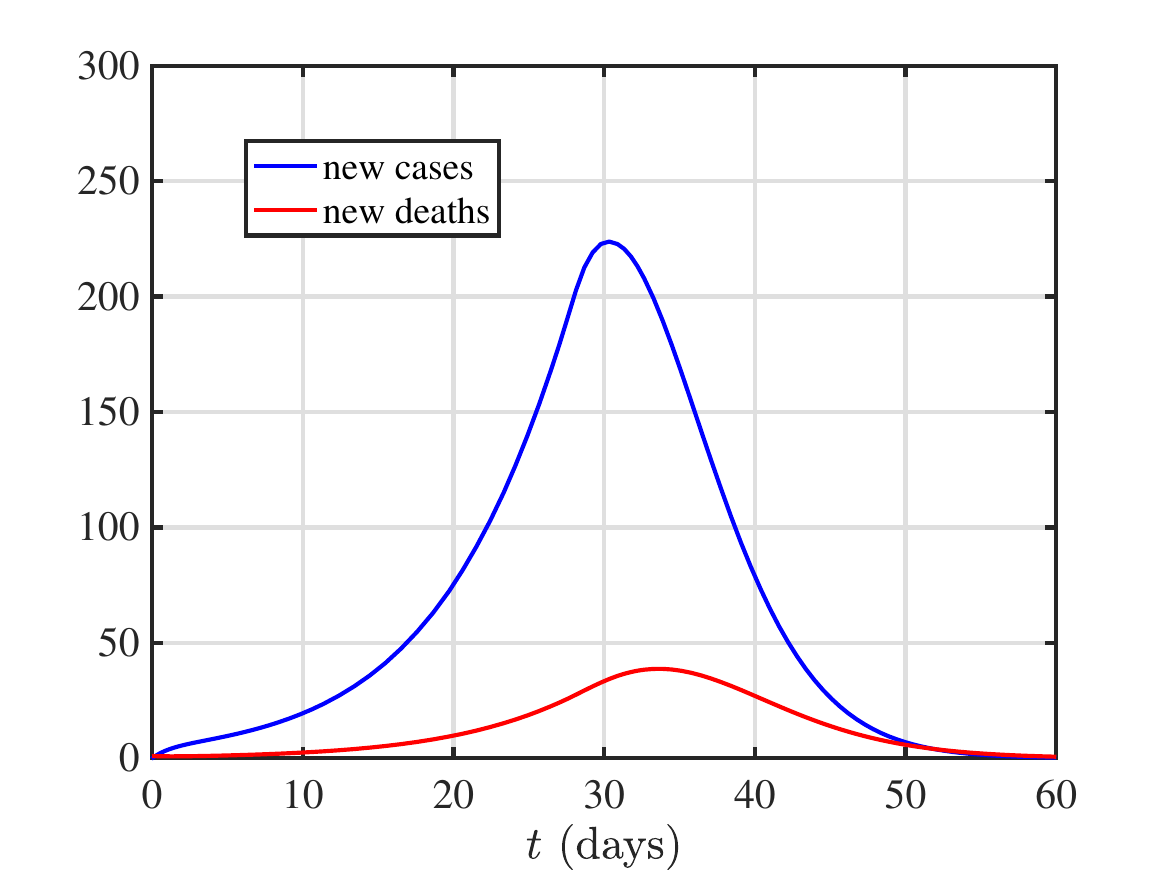} }
  \subfigure[\ SEIR solutions controls removed]
  { \label{f:SEIRPlt-e}
    \includegraphics[width=0.45\columnwidth]{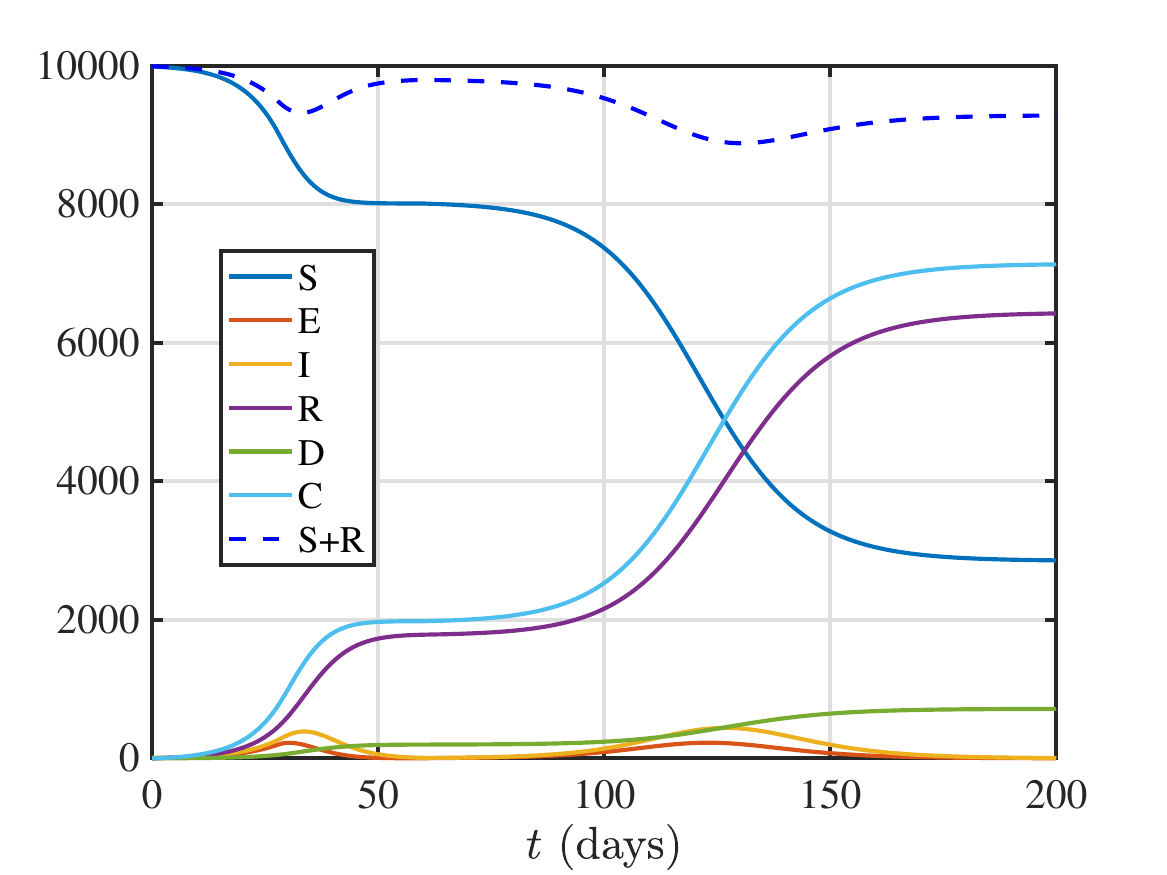} }
  \subfigure[\ SEIR new cases and deaths controls removed]
  { \label{f:SEIRPlt-f}
    \includegraphics[width=0.45\columnwidth]{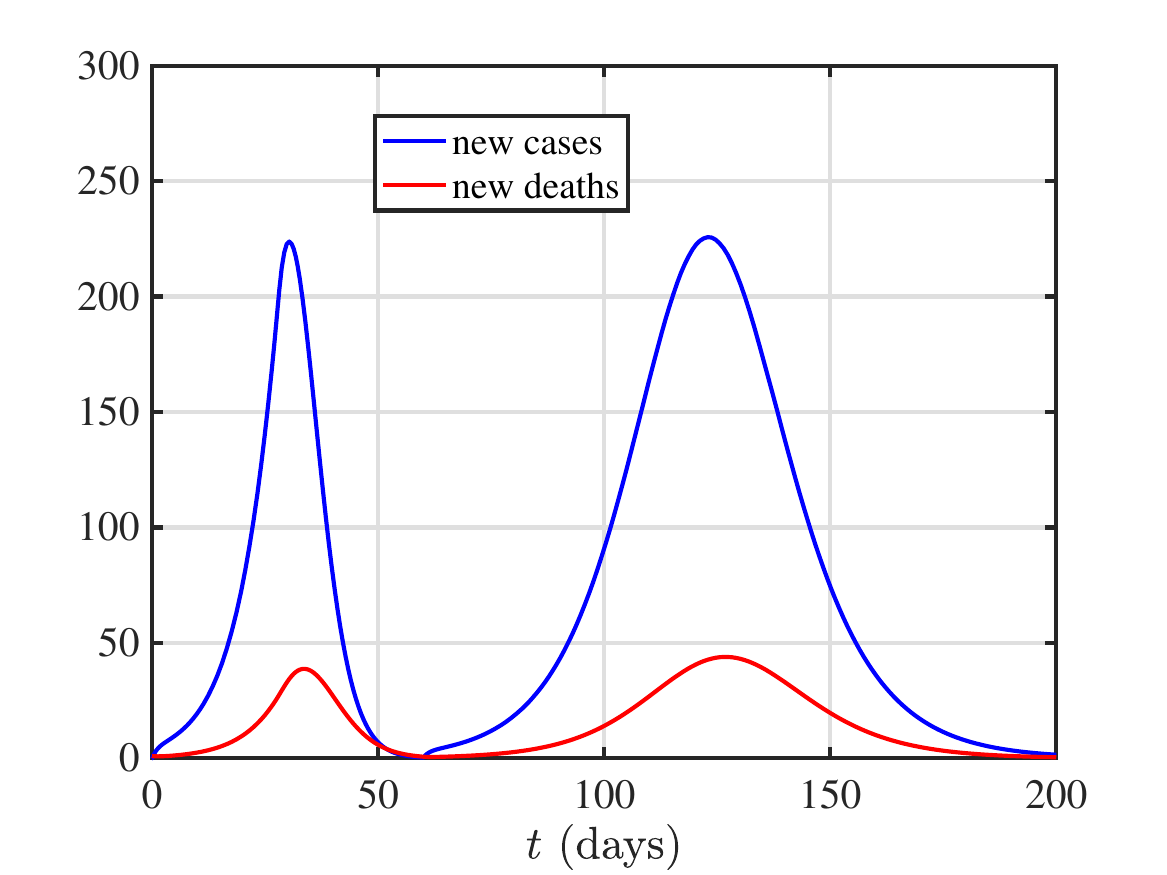} }
  \caption{\label{f:SEIRPlt}Plots of solutions of the SEIR equations
  \ef{e:SEIR3} showing the effect of social distancing,  
  for the case when $\beta_0 = 1/2$, $\sigma = 1/2$, and $\gamma = 1/4$, 
  with the initial values $N_0 = 10000$, $S_0 = N_0 - I_0$,
  $E_0 = 0$, $I_0 = 10$, $R_0 = D_0 = C_0 = 0$, and $f = 0.10$.
  Here $\alpha_0 = 1/8$ and $t_c = 28$ for plots (c) and (d)
  and controls removed at $t = 60$ for plots (e) and (f).}
\end{figure}

One can simulate control measures by allowing the transmission rate $\beta$ to be a function of time, in which case \ef{e:SEIRbasic} becomes
\begin{subequations}\label{e:SEIR3}
\begin{align}
   \dv{S}{t}
   &=
   - \beta(t) \, S(t) I(t) / N \>,
   \label{e:SEIR3-a} \\
   \dv{E}{t}
   &=
   \beta(t) \, S(t) I(t) / N - \sigma \, E(t) \>,
   \label{e:SEIR3-b} \\
   \dv{I}{t}
   &=
   \sigma \, E(t) - \gamma \, I(t) \>,
   \label{e:SEIR3-c} \\
   \dv{R}{t}
   &=
   f \, \gamma \, I(t)
   \qc
   \dv{D}{t}
   =
   (1 - f) \, \gamma \, I(t)
   \qc
   \dv{C}{t}
   =
   \sigma \, E(t) \>.
   \label{e:SEIR3-d}   
\end{align}
\end{subequations}
The effect of social distancing can be studied by allowing the transmission rate $\beta(t)$ to remain constant up to a time $t_0$, at which time control measures are gradually put in place that exponentially reduce the rate.  Choosing $\beta(t)$ of the form, 
\begin{equation}\label{e:control1}
   \beta(t)
   =
   \beta_0 
   \begin{cases}
      1 & \text{for $t < t_0$,} \\
      \exp[ - \alpha_0 \, ( t - t_0 )\,] & \text{for $t > t_0$,}
   \end{cases} 
\end{equation}
we show the results of this calculation in Fig.~\ref{f:SEIRPlt}.  Here we have taken  
$\beta_0 = 1/2$, $\sigma = 1/2$, $\gamma = 1/4$, and $f = 0.10$, and with initial values of  $N_0 = 10000$, $S_0 = N_0 - I_0$, $I_0 = 10$, $E_0 = R_0 = D_0 = C_0 = 0$.
In Figs.~\ref{f:SEIRPlt-a} and \ref{f:SEIRPlt-b} no control measures were in place.  In Figs.~\ref{f:SEIRPlt-c} and \ref{f:SEIRPlt-d} control measures were set in place at $t_c = 28$ days with $\alpha_0 = 1/8$, where we see a dramatic decrease in total number of infections and deaths.  Figs.~\ref{f:SEIRPlt-e} and \ref{f:SEIRPlt-f} shows what happens in this case when control measures are (prematurely) removed at $t = 60$ days when it seems that the number of infectious cases has dropped considerably, but since this is a very infectious disease, the contagious effect of a very small number of cases is still very much evident.  At about $t = 120$ days a considerable increase in the number of infected cases takes place and there is a huge secondary spike in the number of new cases and deaths shown in Fig.~\ref{f:SEIRPlt-f}.  The total number of resulting deaths remains the same as without controls at all, but the length of time of the epidemic has stretched out considerable.  This dramatically shows the consequences of removing controls too soon.

%
%
%
%

Various additions and modification of the SEIR model have been studied by by Berezovskaya \cite{r:Berezovskaya:2005fd} and Wang, \etal\ \cite{WANG20122240}.  See also Brauer \cite{doi:10.1080/17513758.2018.1469792}, Diekmann \cite{doi:10.1098/rsif.2009.0386}, and Miller \cite{Miller:2012aa}.  

%
%
\subsubsection{\label{sss:machine}Machine learning models}

The model of Dandekar and Barbastathis \cite{Dandekar2020.04.03.20052084} (referred to as the MIT model) uses a neural network to find the best fit for a quarantine function.  They start with a simplified SIR model, eliminating the incubation period state $E$, so as to reduce the number of parameters.  Instead of allowing the reproduction rate $\calR$ to become a function of time, they introduce a quarantine fraction $Q(t) = \gamma \, q(t)$ and a quarantine population $T(t) = Q(t) \, I(t) = \gamma \, q(t) \, I(t) $, so that the SIR model equations become:
\begin{subequations}\label{e:MIT}
\begin{align}
   \dv{S}{t}
   &=
   - \beta \, S(t) I(t) / N \>,
   \label{e:MIT-a} \\
   \dv{I}{t}
   &=
   \beta \, S(t) I(t) / N - \gamma \, [ 1 + q(t) ] \, I(t) \>,
   \label{e:MIT-b} \\
   \dv{R}{t}
   &=
   \gamma \, I(t) \>,
   \label{e:SEIbasic-c} \\
   \dv{T}{t}
   &=
   \gamma \, q(t) \, I(t) \>.
\end{align}
\end{subequations}
In dimensionless time $\tau = \gamma \, t$, these SIR equations are described by a single parameter $\calR_0 = \beta/\gamma$ and an unknown quarantine function $q(t)$.    The problem is then to find the best quarantine function, which they implement by using machine learning with a neural network.  That is they set $Q(t) = \mathrm{NN}[\, W, U(t) \,]$, where $U(t) = \{\, S(t),I(t),R(t),T(t) \,\}$ and $\mathrm{NN}$ is a neural network consisting of an activation function which connects all the functions represented by $U(t)$ at time $t$ to the quarantine fraction $Q(t)$, which is in itself a solution of \ef{e:MIT}.  They use a neural network consisting of a ``2-layer densely connected network with 10 units in the hidden layer and the ReLU activation function.'' The network results in 64 weights $W_i$ which are tuned to obtain the best fit to the data.  The neural network is used to both solve the model differential-algebraic equation (DAE) system and the error loss by a local sensitivity analysis to the data (See Cao \etal\ \cite{doi:10.1137/S1064827501380630} and Rackauckas \etal\ \cite{rackauckas2020universal}).  This is a very sophisticated calculation using the {\tt JULIA} computer language, developed at MIT.   
(See Rackauckas \etal\ \cite{Rackauckas:2019aa} and the {\tt JULIA} website \url{https://julialang.org/} for further details.)

%
%
\subsection{\label{ss:geomods}Geographical models}

\begin{figure}[t]
  \centering
  \subfigure[\ $\nu = f = 0$.]
     {\label{f:denplt}
        \includegraphics[width=0.45\columnwidth]{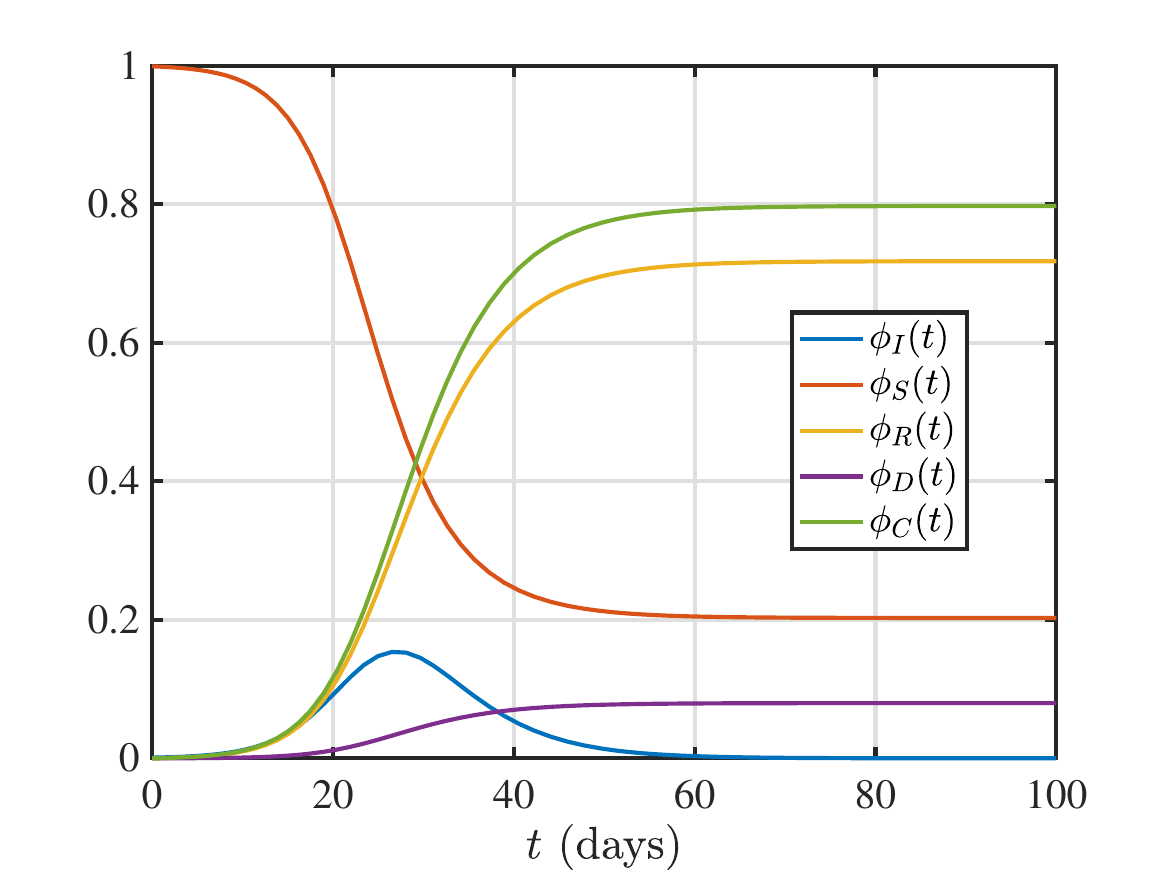} }
  \subfigure[\ $\nu = f = 1/100$]
     {\label{f:denpltosc}
        \includegraphics[width=0.45\columnwidth]{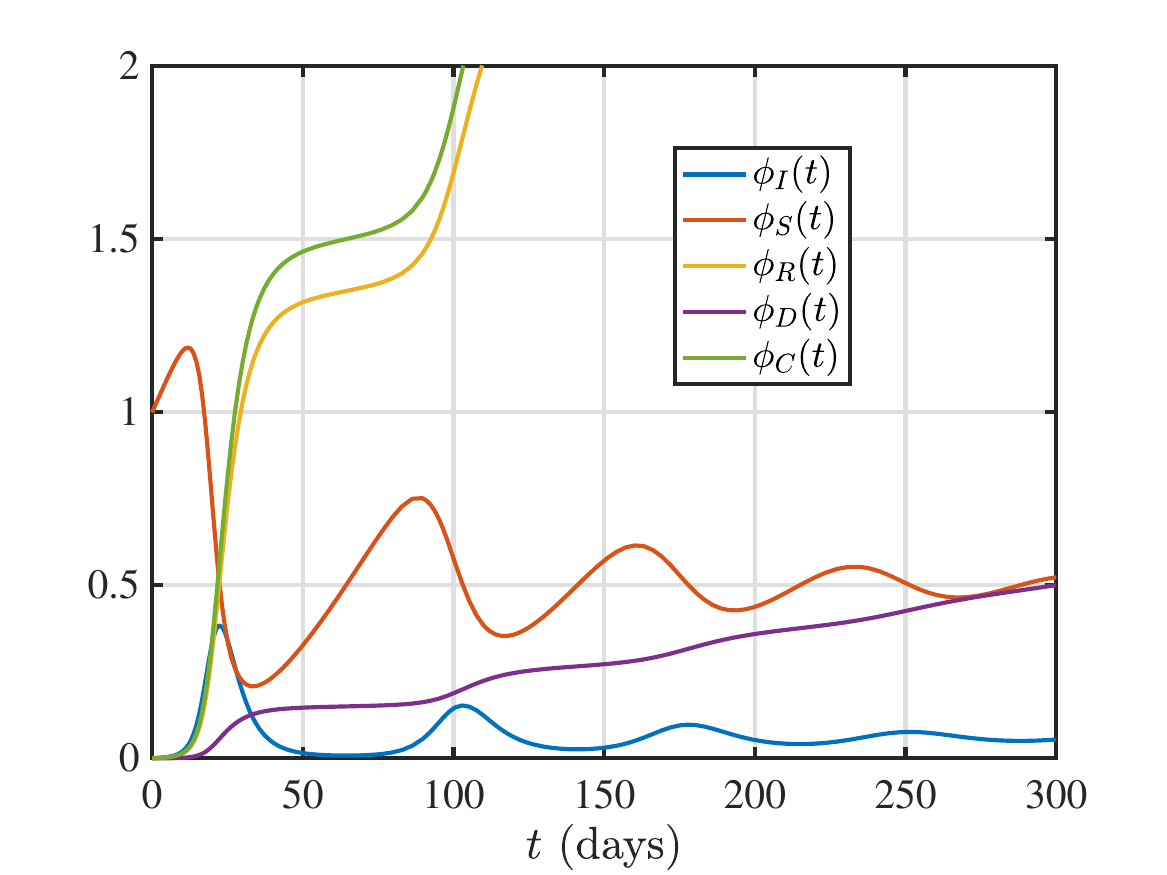} }
  \caption{\label{f:States} Solutions of \ef{e:SIRdensities} for
  $\lambda = 1/2$, $\mu = 1/4$, $g = 0.10$.  Panel (a) shows the
  results when $\nu = f = 0$.  Panel (b) shows the results when
  $\nu = f = 1/100$. }
\end{figure}

The extension of SEIR-type models to include movement of individuals in a geographical region is straightforward.  However to simplify the dynamics, we will use the simpler SIR model described by the reactions,
\begin{equation}\label{e:SIRreactions}
   S + I \xrightarrow{k_1} 2 \, I
   \qc
   I \xrightarrow{k_2} 0 
   \qc
   S \xrightarrow{k_3} 0
   \qc
   0 \xrightarrow{k_4} S
   \>.
\end{equation}
Here we have allowed for a death (or recovery) rate $k_2$ for the infected population and death $k_3$ and 
birth $k_4$ rates for the susceptible population.  Dead or recovered individuals do not enter into the dynamics.  The $d$-dimensional region is divided into cells labeled by the index $i$, so that populations are now labeled by the set $\{\, S_{i}(t),I_{i}(t) \, \}$.  The susceptible and infected populations are allowed to jump into neighborhood regions with rates $d_S$ and $d_I$ respectively,
\begin{equation}\label{E:jumping}
   I_{i}(t)
   \xrightarrow{d_I}
   I_{i\pm 1}(t)
   \qc
   S_{i}(t)
   \xrightarrow{d_S}
   S_{i\pm 1}(t)
   \>.
\end{equation}
Neighboring regions can be as far away as New York and Florida, as these regions are connected by exchanging populations.  We assume that the recovered are immune.  Passing over to a continuum density description, we set
\begin{equation*}\label{e:densities}
   \phi_S(\vb{x},t) = S_{i}(t)/L^d
   \qc
   \phi_I(\vb{x},t) = I_{i}(t)/L^d \>.
\end{equation*}
The total population density is now given by $\rho_0 = N / L^d$.  
Then the continuum version of the SIR equations \ef{e:SEIRbasic} becomes:
\begin{subequations}\label{e:SIRdensities}
\begin{align}
   \partial_t\phi_I(\vb{x},t)
   &=
   [\, D_I \laplacian - \mu \,] \, \phi_I(\vb{x},t)
   +
   \lambda \, \phi_S(\vb{x},t) \, \phi_I(\vb{x},t) \>,
   \label{e:SEIRcon-a} \\
   \partial_t\phi_S(\vb{x},t)
   &=
   [\, D_S \laplacian - \nu \,] \, \phi_S(\vb{x},t)
   - 
   \lambda \, \phi_S(\vb{x},t) \, \phi_I(\vb{x},t)
   +
   f \>,
   \label{e:SEIRcon-b}
\end{align}
\end{subequations}
Here $\lambda = k_1 L^d = \beta / \rho_0$, $\mu = k_2 L^d$, $\nu = k_3 L^d$, $f = k_4$, $D_S = 2 d_S/h^2$, and $D_I = 2 d_I/h^2$.  In this model, dispersion takes place as a random walk.  We recognize Eqs.~\ef{e:SIRdensities} as being a set of coupled non-linear \Schrodinger\ equations in imaginary time, but with \emph{real} ``wave functions.''  We can add recovered, dead, and accumulated infection density by the equations,
\begin{subequations}\label{e:RDCdensities}
\begin{align}
   \partial_t\phi_R(\vb{x},t)
   &=
   (1 - g) \, \mu \, \phi_I(\vb{x},t) \>,
   \label{e:Rdensity} \\
   \partial_t\phi_D(\vb{x},t)
   &=
   g \, \mu \, \phi_I(\vb{x},t) \>,
   \label{e:Ddensity} \\
   \partial_t\phi_C(\vb{x},t)
   &=
   \mu \, \phi_I(\vb{x},t) \>,
   \label{e:Cdensity}   
\end{align}
\end{subequations}
which do not enter into the dynamics.  Here $g$ is the fraction of infected individuals that die.  
In Fig.~\ref{f:denplt} we show solutions of \ef{e:SIRdensities} for the homogeneous case when $\lambda = 1/2$, $\mu = 1/4$, $g = 0.10$ and have set $\nu = f = 0$.  Compare this figure with the results in Fig.~\ref{f:SEIRPlt-a} using the SEIR equations.  Fig.~\ref{f:denpltosc} shows the results when the birth and death rate of the susceptible population is set to $\nu = f = 1/100$, showing an oscillation in the densities surviving for 200 days.  This type of oscillatory behavior is discussed in Section~\ref{sss:temporal} below. 

In Fig.~\ref{f:DenX}, we show solutions of Eqs.~\ef{e:SIRdensities} in one physical dimension for the
case when $\lambda = 1/2$, $\mu = 1/4$, $g = 0.10$, $\nu = f = 1/100$, and with $D_S = 10$ and $D_I = 2$.
Here $N_0 = 100$ individuals and $L = 100$ miles.  

We numerically solved Eqs.~\ef{e:SIRdensities} using the exponential Runge-Kutta fourth order ETDRK4 algorithm scheme of Cox and Matthews \cite{COX2002430} patterned after MatLab codes by Kassam \cite{r:Kassam:2003vn,r:Kassam:2005wy}. In the code, we used aliasing with a 2/3 rule and enforced real densities.  In Fig.~\ref{f:DenX}, we show solutions in one physical dimension for the
case when $\lambda = 1/2$, $\mu = 1/4$, $g = 0.10$, $\nu = f = 1/100$, and with $D_S = 10$ and $D_I = 2$.
Here $N_0 = 100$ individuals and $L = 100$ miles.  We started out the dynamics with a Gaussian distribution of 5 infected individuals at the origin and 95 susceptible individuals distributed evenly over the region.  In this case, susceptible individuals make a random walk of about 4.4 miles per day, infected individuals about 2 miles per day.  The results show a steady progression of the infection out from the origin until the region is filled with the blue steady state solution.  The birth and death rate of the susceptible population sustains a steady recovery and death rate after 50 days, not a very good outcome.  No Turing patterns are formed in this scenario.

We examine homogeneous and steady state solutions of this model in the next section.
 
%
%
%
%

%
%
\begin{figure}[t]
  \centering
  \subfigure[\ $S(t)$, $I(t)$, $R(t)$ and $D(t)$.]
     {\label{f:denX-a}
        \includegraphics[width=0.45\columnwidth]{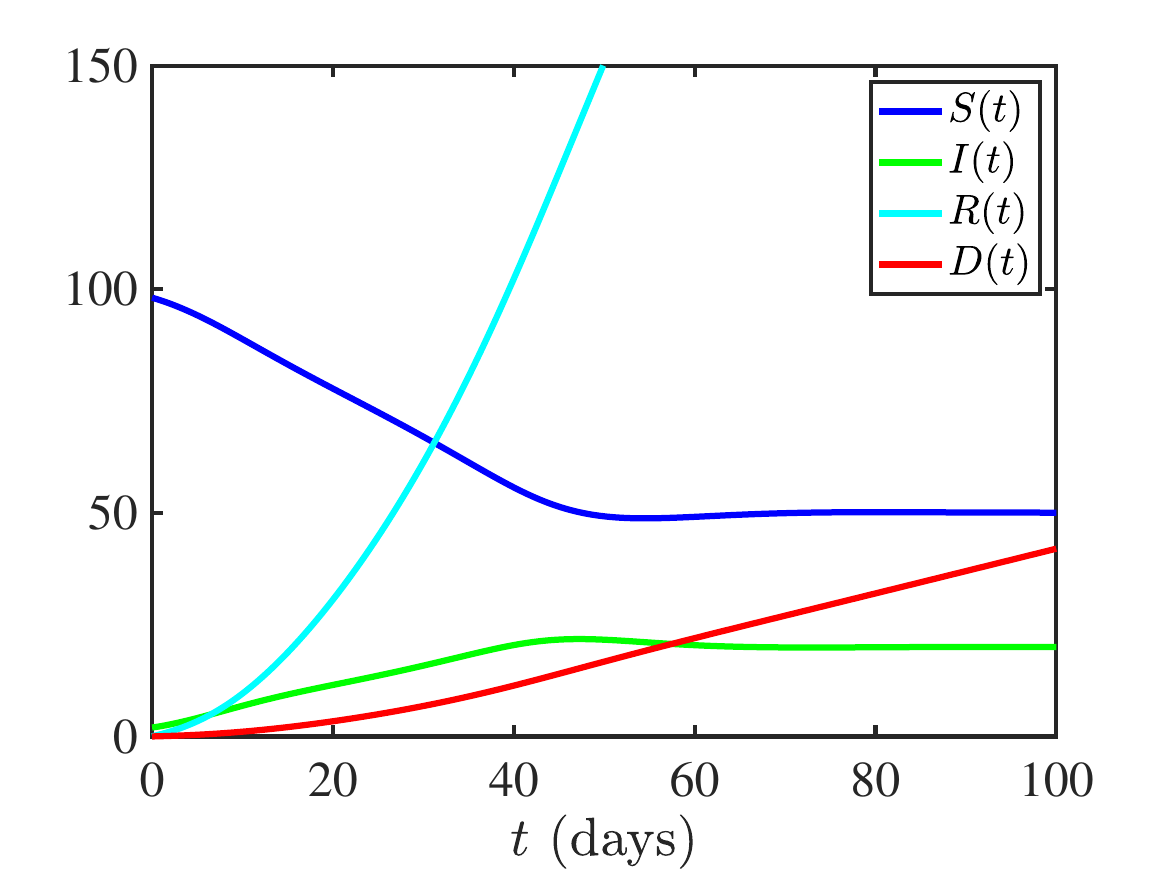} }
  \subfigure[\ Solutions at $T = 35$ days]
     {\label{f:denX-b}
        \includegraphics[width=0.45\columnwidth]{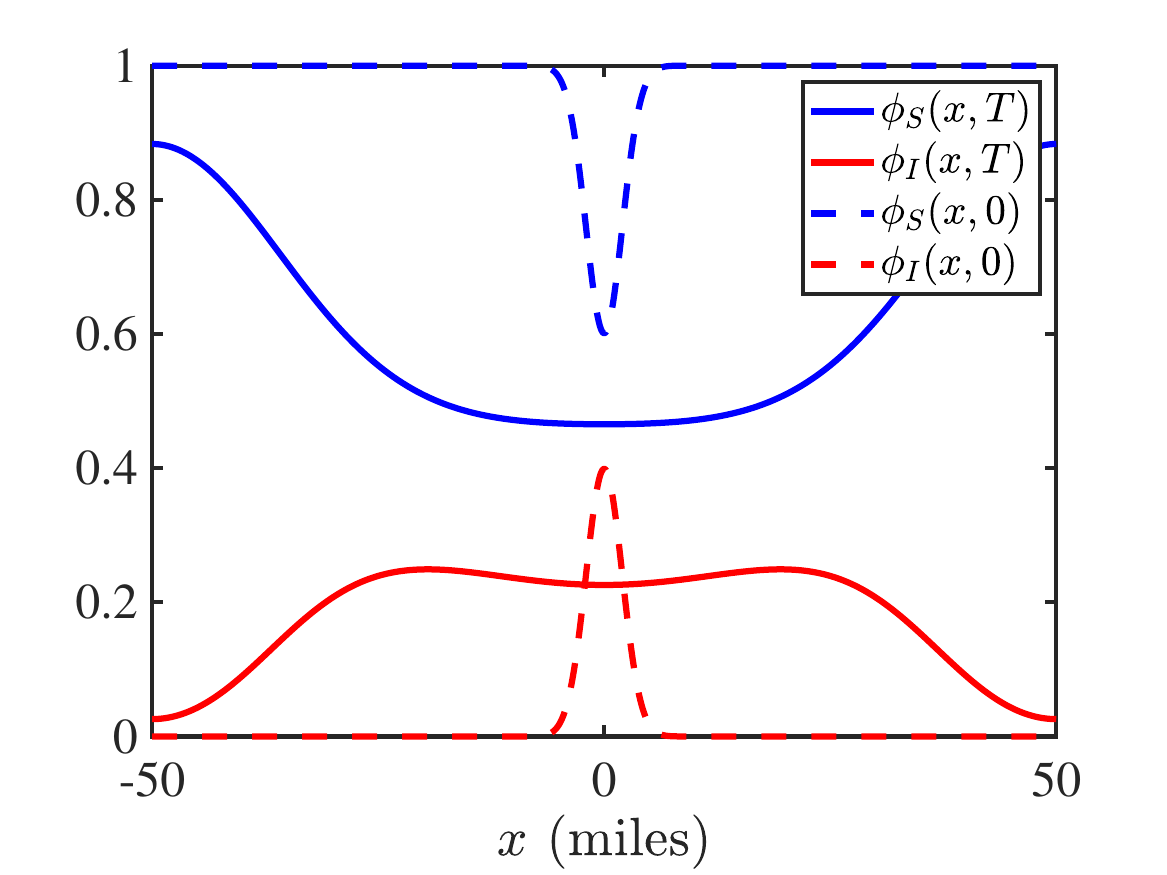} }
  \subfigure[\ $\phi_S(x,t)$ and $\phi_I(x,t)$]
     {\label{f:denX-c}
        \includegraphics[width=0.45\columnwidth]{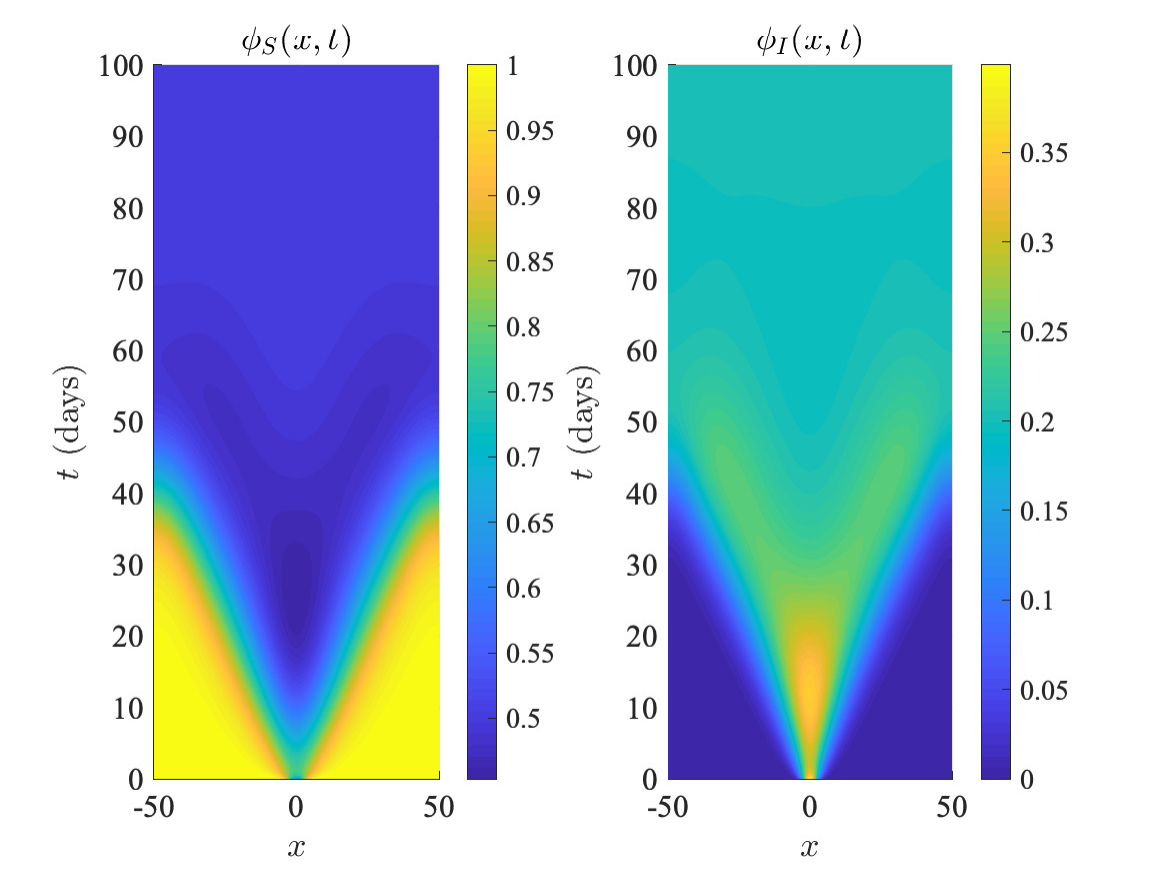} }
  \caption{\label{f:DenX} Solutions of \ef{e:SIRdensities} for
  $\lambda = 1/2$, $\mu = 1/4$, $g = 0.10$, $\nu = f = 1/100$,
  and with $D_S = 10$ and $D_I = 2$.
  Panel (a) shows the integrated results, panel (b) the results
  after $T = 35$ days, and panel (c) a density plot of the dynamics.
  Here $N_0 = 100$ individuals and $L = 100$ miles.}
\end{figure}
%
%

%
%
\subsubsection{\label{ss:HSS}Homogeneous and steady state solutions}

\begin{figure}[t]
  \centering
  \subfigure[\ $\phi_I^{(B)}$ (in red) and $\phi_S^{(B)}$ (in blue)]
     {\label{f:ISbluestates}
        \includegraphics[width=0.45\columnwidth]{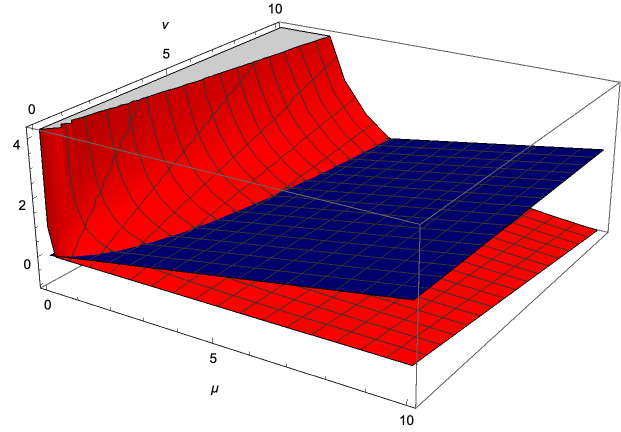} }
  \subfigure[\ Turing curves]
     {\label{f:Turing}
        \includegraphics[width=0.45\columnwidth]{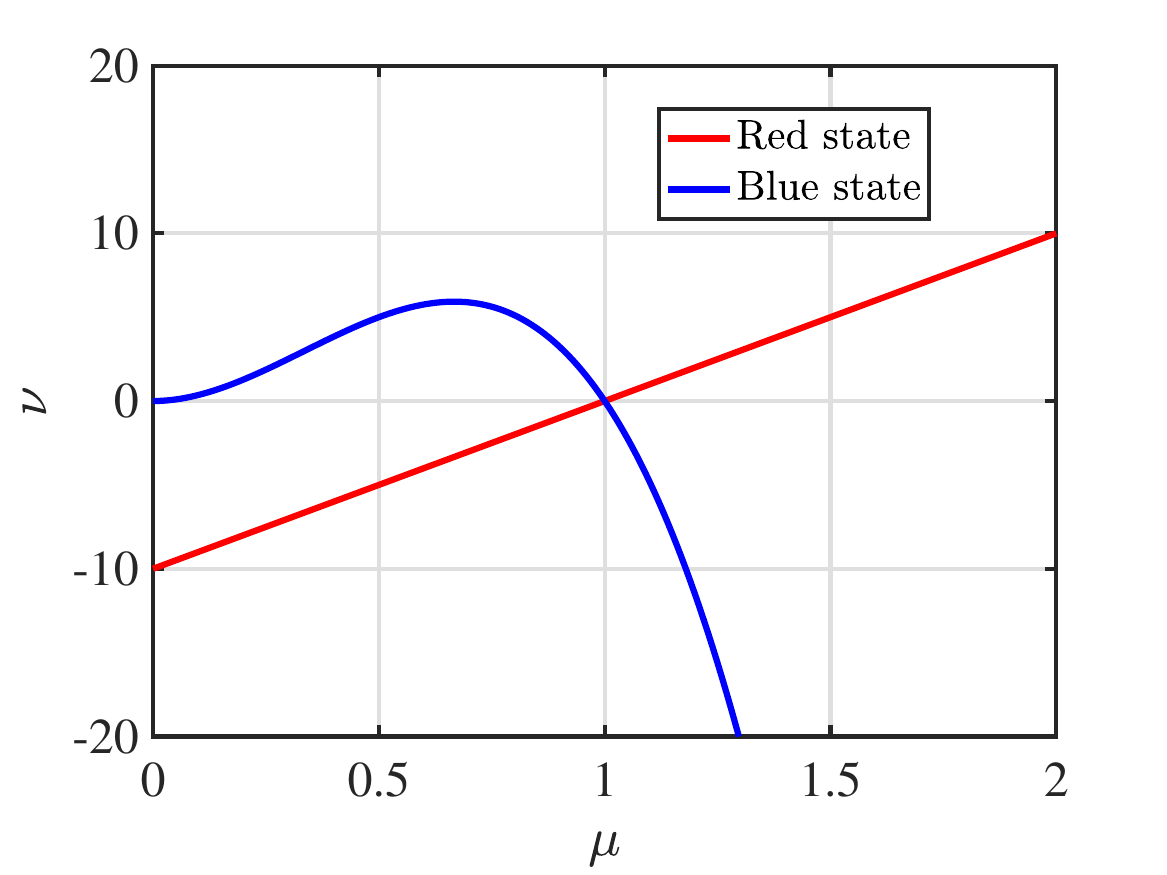} }
  \caption{\label{f:States} (a) Plots of $\phi_I^{(B)}$ (in red) 
  and $\phi_S^{(B)}$ (in blue) as a function of $\mu$ and $\nu$
  for $f = \nu$ and $\lambda = 5$.  At the intersection of the two
  surfaces, the populations are equal. (b) Plots of the Turing
  curves from Eqs.~\ef{e:RedTuring1} and \ef{e:bluecase2} for 
  the case when $\lambda = 1$ and $D_I/D_S = 0.1$.}
\end{figure}

Ignoring the recovered population, homogeneous and steady state solutions exist for the rest of the SIR model.  Setting $\phi_{\alpha}(x,t) = \phi_{\alpha}$, they are given by solutions of the equations:
\begin{subequations}\label{e:HSS}
\begin{align}
   \nu \, \phi_{I} - \lambda \, \phi_{S} \, \phi_{I}
   &=
   0 \>.
   \label{e:HSS-a} \\
   \nu \, \phi_{S} + \lambda \, \phi_{S} \, \phi_{I} 
   &=
   f \>,
   \label{e:HSS-b}
\end{align}
\end{subequations}
There are two sets of solutions of \ef{e:HSS} given by
\begin{subequations}\label{e:RedBlue}
\begin{gather}
   \phi_I^{(R)} = 0 \qc \phi_S^{(R)} = \frac{f}{\nu} \>,
   \label{e:Red} \\
   \phi_I^{(B)} = \frac{f}{\mu} - \frac{\nu}{\lambda} \qc
   \phi_S^{(B)} = \frac{\mu}{\lambda} \>.
   \label{e:Blue}
\end{gather}
\end{subequations}
Solutions \eqref{e:Red} are labeled ``red,'' and those in \eqref{e:Blue} ``blue.''  The red solutions consist of a uniform population of susceptible individuals and no infected individuals, which is obviously stable.  It is usual to set the birth and death rates of susceptible individuals numerically equal, in which case $f = \nu$ and $\phi_S^{(R)} = 1$.  It is also useful to set $\mu = \nu + \kappa$, with $\kappa > 0$.  With this notation, blue solutions are possible if
\begin{equation}\label{e:BlueReq}
   \frac{f}{\mu} - \frac{\nu}{\lambda} 
   =
   \frac{\nu}{\mu \lambda} \, (\lambda - \mu )
   =
   \frac{\nu \, ( \lambda - \nu - \kappa )}{\lambda \, (\nu + \kappa) }
   \ge 
   0 \>.
\end{equation}
In this case, both infected and healthy individuals can coexist.  A plot of $\phi_I^{(B)}$ (in red) and $\phi_S^{(B)}$ (in blue) as a function of $\kappa$ and $\nu$ is shown in Fig.~\ref{f:ISbluestates}.  In some regions $\phi_I^{(B)}$ is greater than $\phi_S^{(B)}$.
%
%
%
%

%
%
\subsubsection{\label{sss:stability}Stability analysis}

Let us define
\begin{equation}\label{e:Phidefs}
   \Phi(x,t)
   =
   \begin{pmatrix}
      \phi_I(x,t) \\
      \phi_S(x,t)
   \end{pmatrix}
   \qc
   \Phi_0
   =
   \begin{pmatrix}
      \phi_I \\
      \phi_S
   \end{pmatrix}
   \qc
   \delta \Phi
   =
   \begin{pmatrix}
      \delta \phi_I \\
      \delta \phi_S
   \end{pmatrix} \>,
\end{equation}
where $\Phi_0$ are solutions of the homogeneous and steady state solutions \ef{e:HSS}.  Then the stability of the solutions of the rate equations can be studied by setting
\begin{equation}\label{e:Phiexpand}
   \Phi(x,t)
   =
   \Phi_0
   +
   \delta \Phi \, \rme^{\rmi ( k x - \omega t) } \>,
\end{equation}
and expanding the result to first order in $\delta \Phi$.  This gives an equation of the form,
\begin{subequations}\label{e:PhiGinvdef}
\begin{gather}
   G^{-1}_{k,\omega}[\, \Phi_0 \,] \, \delta \Phi
   = 0 \>,
   \label{e:PhiGinvdef-a} \\
   G^{-1}_{k,\omega}[\, \Phi_0 \,]
   =
   \begin{pmatrix}
      -\rmi \omega + D_I k^2 + \mu - \lambda \, \phi_S \>, & - \lambda \, \phi_I \\
      \lambda \, \phi_S \>, & -\rmi \omega + D_S k^2 + \nu + \lambda \, \phi_I
   \end{pmatrix} \>.
   \label{e:PhiGinvdef-b}
\end{gather}
\end{subequations}
Non-trivial solutions are possible if
\begin{align}\label{e:detM}
   \det{ G^{-1}_{k,\omega}[\, \Phi_0 \,] }
   =
   - \omega^2 - \rmi \, B_k \, \omega + C_k
   =
   -
   (\omega - \omega_k^{(+)}) \, (\omega - \omega_k^{(-)})
   =
   0 \>.
\end{align}
where
\begin{subequations}\label{e:BC}
\begin{align}
   B_k
   &=
   (D_I + D_S) \, k^2
   +
   \lambda \, ( \phi_I - \phi_S )
   +
   \mu + \nu \>,
   \label{e:B} \\
   C_k
   &=
   D_I D_S \, k^4
   +
   [\, 
      D_I ( \nu + \lambda \phi_I )
      +
      D_S ( \mu - \lambda \phi_S ) \,
   ] \, k^2
   +
   \lambda ( \mu \phi_I - \nu \phi_S )
   +
   \mu \nu \>.
   \label{e:C}
\end{align}
\end{subequations}
and
\begin{equation}\label{e:eigenvalues}
   \omega_k^{(\pm)}
   =
   \rmi \, \frac{B_k}{2} 
   \pm
   \rm i \, \sqrt{ \Bigl ( \frac{B_k}{2} \Bigr )^2 - C_k } \>.
\end{equation}

%
%
\subsubsection{\label{sss:temporal}Temporal stability}

Homogeneous and oscillatory solutions are found for $k=0$ when $B_0 = 0$ and $C_0 < 0$. 
From \ef{e:BC}, this requires
\begin{subequations}\label{e:BCtemp}
\begin{align}
   \nu + \mu + \lambda \, (\, \phi_I - \phi_S \,)
   =
   0 \>,
   \label{e:Btemp} \\
   \mu \, \nu + \lambda \, ( \mu \, \phi_I - \nu \, \phi_S )
   \le 0 \>.
   \label{e:Ctemp}
\end{align}
\end{subequations}
For the red state, $\phi_S = f/\nu$ and $\phi_I = 0$, so that \ef{e:Btemp} required that $\lambda f/\nu= \nu + \mu$, in which case $C_0 = \nu ( \mu - \lambda ) = - \nu^2 \le 0$ is always satisfied so that the red states are temporaly stable.  The oscillation frequencies are $\omega_0^{(\pm)} = \pm \nu$.  The blue state is unstable for all values of the parameters.  This temporal instability is called Hopf bifurcation.  
This oscillation is illustrated in Fig.~\ref{f:denpltosc} where the simulation was started in the Red state where $\phi_S = 1$ and $\phi_I = 0$, with $\nu = f = 1/100$, but winds up asymptotically in the Blue state with $\phi_S = 0.5$ and $\phi_I = 0.02$, which is stable.  Note that the oscillation period is approximately $1/\nu = 100$ days.  In order to have oscillations $\nu$ and $f$ must be non-zero --- that is oscillations require a constant entering and leaving of the physical region, which is probably to be expected in any epidemic.  The rate of the entering and leaving govern the the oscillation period.  Note the steady increase in deaths in this scenario.   

%
%
\subsubsection{\label{sss:turing}Spacial stability}

Inhomogeneous and stable steady state patterns can be set up if certain conditions are met.  These patterns are called Turing patterns --- the condition under which the patters emerge are called Turing bifurcations.  For patterns to emerge, we must have $C_k \le 0$.  The critical value is when $C_{k_c} = 0$, or from \ef{e:C} when
\begin{equation}\label{e:C-critical}
   D_I D_S \, k_c^4 
   +
   [\, D_I (\nu + \lambda \, \phi_I) + D_S (\mu - \lambda \, \phi_S ) \, ] \, k_c^2
   +
   \lambda \, ( \mu \phi_I - \nu \phi_S ) + \mu \nu
   =
   0 \>.   
\end{equation}
As a function of $k^2$, $C_k$ is a parabolic curve with positive curvature,
\begin{subequations}\label{e:Cparabolic}
\begin{align}
   \pdv{C_k}{k^2} \Big |_{k_c}
   &=
   2 \, D_I D_S \, k_c^2
   +
   [\, D_I (\nu + \lambda \, \phi_I) + D_S (\mu - \lambda \, \phi_S ) \, ] \>,
   \label{e:Cparabolic-a} \\
   \pdv[2]{C_k}{(k^2)}  \Big |_{k_c}
   &=
   2 \, D_I D_S > 0 \>.  
   \label{e:Cparabolic-b}
\end{align}
\end{subequations}
The minimum of the curve is at
\begin{equation}\label{e:Cmin}
    k_c^2
    =
    -
    \frac{ D_I (\nu + \lambda \, \phi_I) + D_S (\mu - \lambda \, \phi_S ) }
         { 2 \, D_I D_S } \>.
\end{equation}
Substituting this into \ef{e:C-critical} gives:
\begin{align}\label{e:Critical1}
   [\, D_I (\nu + \lambda \, \phi_I) + D_S (\mu - \lambda \, \phi_S ) \, ]^2 
   &=
   4 \, D_I D_S \, 
   [\, \mu \, \nu + \lambda \, ( \mu \, \phi_I - \nu \, \phi_S ) \,]
   \\
   &
   \hspace{-4em}
   =
   4 \, D_I (\nu + \lambda \, \phi_I) \, D_S (\mu - \lambda \, \phi_S )
   -
   4 \, D_I D_S \, \lambda^2 \, \phi_I \, \phi_S \>.
   \notag
\end{align}
Choosing the negative sign for the square root and substitution into \ef{e:Cmin}, the critical wave number is given by
\begin{equation}\label{e:kcvalue}
   k_c
   =
   \sqrt{ 
      \frac{ \mu \, \nu + \lambda \, ( \mu \, \phi_I - \nu \, \phi_S ) }
           { 4 \, D_I D_S } }
\end{equation}
Simplifying \ef{e:Critical1} gives
\begin{align}\label{e:Critical2}
   -
   D_I^2 (\nu + \lambda \, \phi_I)^2
   +
   &2 \, D_I (\nu + \lambda \, \phi_I) \, D_S (\mu - \lambda \, \phi_S )
   -
   D_S^2 (\mu - \lambda \, \phi_S )^2
   \\
   &
   =
   4 \, D_I D_S \, \lambda^2 \, \phi_I \, \phi_S \>,
   \notag
\end{align}
which can be written as
\begin{equation}\label{e:Critical3}
   [\, D_I (\nu + \lambda \, \phi_I) - D_S (\mu - \lambda \, \phi_S ) \, ]^2
   =
   4 \, D_I D_S \, \lambda^2 \, \phi_I \, \phi_S \>.
\end{equation}
For the red state, \ef{e:Critical3} becomes:
\begin{equation}\label{e:RedTuring1}
   D_I \, \nu = D_S (\mu - \lambda ) 
   \qq{$\Rightarrow$}
   \frac{\mu - \lambda}{\nu}
   =
   \frac{D_I}{D_S} \>,
\end{equation}
which defines the Turing line.  For the red state, from \ef{e:kcvalue} along this line the critical wave number is given by
\begin{equation}\label{e:kcTuringRed}
   k_c
   =
   \sqrt{ \frac{ \nu\, ( \mu - \lambda )}{4 \, D_I D_S} }
   =
   \frac{\nu}{2 D_S} \>.
\end{equation}
For the blue state,
\begin{equation}\label{e:bluecase1}
   \nu + \lambda \, \phi_I = \frac{\nu \lambda}{\mu}
   \qc
   \lambda \, \phi_S = \mu \>.
\end{equation}
Subsituting this into \ef{e:Critical2} gives the condition,
\begin{equation}\label{e:bluecase2}
   4 \, \frac{ \mu^2 }{ \lambda^2 } \, 
   \Bigl ( \frac{\lambda - \mu}{\nu} \Bigr )
   =
   \frac{D_I}{D_S} \>.
\end{equation}
The Turing curves are shown in Fig.~\ref{f:Turing}.
Both of these states can exhibit Turing patterns.  

%
%
\subsection{\label{ss:stochastic}Stochastic models}

Stochastic models of processes that lead to rate equations like \ef{e:SIRdensities} start from developing a master equation for the microscopic reaction kinematics.  Assuming that the processes are Markovian, one can either solve the master equation numerically for population number using an algorithm, such as the Gillespie algorithm, for each process and reaction selected at random, or develop rate equations for the densities, called \emph{Langevin equations}, which includes internal noise generated by the microscopic model.  Details of this derivation are given in Appendix~\ref{s:Langevin}. The Gillespie algorithm is discussed in Appendix~\ref{s:micro}.

%
%
\subsubsection{\label{sss:Langevin}Langevin equations}

In appendix~\ref{s:Langevin}, we derive Langevin equations for the SIR model which incorporate noise generated internally by the stochastic nature of the microscopic model.  Solving Langevin-type equations are an alternate way to simulating the master equation directly.  For the SIM model, the Langevin equations are given by:
\begin{subequations}\label{e:LangevinI}
\begin{align}
   \bigl (\,
      \partial_t
      - 
      D_E \laplacian 
      +
      \mu \,
   \bigr ) \, \phi_E(x)
   -
   \lambda \, \phi_S(x) \, \phi_I(x)
   &=
   \eta_I(x) \>,
   \label{e:LangevinI-a} \\
   \bigl (\,
      \partial_t
      - 
      D_S \laplacian 
      +
      \nu \,
   \bigr ) \, \phi_S(x)
   +
   \lambda \, \phi_S(x) \, \phi_I(x)
   -
   f
   &=
   \eta_S(x) \>,
   \label{e:LagevinI-b}
\end{align}
\end{subequations}
with noise functions,
\begin{subequations}\label{e:noise}
\begin{align}
   \eta_I(x)
   &=
   \sqrt{2 \sigma(x)} \, \theta_1(x)
   \qc
   \sigma(x) = - \lambda \, \phi_{I}(x) \, \phi_S(x) \>,
   \label{e:noise-a} \\
   \eta_S(x)
   &=
   - \sqrt{\sigma(x)/2} \, 
   [\,\theta_1(x) + \rmi \, \theta_2(x) \,] \>.
   \label{e:noise-b}
\end{align}
\end{subequations}
In Appendix~\ref{s:micro} we derive microscopic equations to simulate directly the SIR model.

%
%
\section{\label{s:applications}Applications}

%
%
\subsection{\label{ss:fitting}Fitting models to data}

The problem of applying SEIR models to epidemics is that data is either unreliable or incomplete.  One would like to be able to predict the course of an epidemic at the beginning of the process so that corrective measures can be taken early on.  This is not always possible.  Ideally modelers would like to have data on the number of cases, the number of deaths, the number of recovered, and if possible the number of infected or hospitalized per day as a function of date.  This data is very difficult to get, even from organizations like the CDC or the WHO which \emph{should} provide it.\footnote{I have found it particularly difficult to obtain data from these organizations.}

Fitting data requires the minimization of a cost function, usually taken to be a least-squares function of the form,
\begin{equation}\label{e:Lfunc}
   L(p)
   =
   \sum_{i=1}^{N_d} \sum_{j=1}^{N_m} | \, y_j(t_i,p) - d_j(i) \, |^2 \>,
\end{equation}
where the sum goes over the $N_d$ data points $d_j(i)$ at $t_i$.  Here $y_i(t,p)$ is a vector of model population dimension $N_m$, satisfying the differential algebraic equations
\begin{equation}\label{e:DAE}
   \dot{y}_j = f_j(y,p,t)
   \qc
   y_j(0,p) = y_{j,0}(p)
   \qc
   j = 1,2,\dotsc,N_m \>,
\end{equation}
and where $p = (\,p_1,p_2,\dotsc,p_{N_p}\,)$ is a vector containing the $N_p$ the parameters in the problem.  Note that the initial conditions $y_{j,0}(p)$ can depend on the parameters also.  Minimization of \ef{e:Lfunc} with solutions of the differential equations \ef{e:DAE} is a classic problem in applied mathematics and analysis.  We describe here a gradient method given in papers by Petzold and co-workers \cite{LI2000131,doi:10.1137/S1064827501380630}.  In this method, we start by expanding \ef{e:Lfunc} to first order about a point $p$,
\begin{equation}\label{e:Lp0expand}
   L(p + \Delta p)
   =
   L(p) + \sum_{k=1}^{N_p} \pdv{L(p)}{p_k} \, \Delta p_k + \dotsb
\end{equation}
Now choosing
\begin{equation}\label{e:Deltap}
   \Delta p_k 
   =
   - h \, \pdv{L(p)}{p_k}
   \qq{$\Rightarrow$}
   L(p + \Delta p) - L(p)
   =
   - h \, \sum_{k=1}^{N_p} \Big | \pdv{L(p)}{p_k} \Bigr |^2 < 0 \>,
\end{equation}
we move the point $p$ to a value which gives a smaller value of $L(p)$.  Here $h$ is the step size.
Differentiating \ef{e:Lfunc} with respect to the parameters and setting the result to zero gives
\begin{equation}\label{e:dLdp}
   \pdv{L(p)}{p_k}
   =
   2 \sum_{i=1}^{N_d} \sum_{j=1}^{N_m} \, [\, y_j(t_i,p) - d_j(i) \,] \, s_{jk}(t_i,p) 
   \qc
   s_{jk}(t,p)
   =
   \pdv{y_j(t,p)}{p_k} \>.
\end{equation}
This necessitates finding the solution of not only the original equations \ef{e:DAE}, but the set of differential equations for for $s_{ij}(t,p)$, given by
\begin{equation}\label{e:sDEQs}
   \dot{s}_{ij}(t,p)
   =
   \pdv{f_i(y,t,p)}{y_k} \, s_{kj}(t,p) + \pdv{f_i(y,t,p)}{p_j}
   \qc
   s_{ij}(0,p)
   =
   \pdv{y_{i,0}(p)}{p_j} \>.
\end{equation}
So we must find the simultaneous solution of Eqs.~\ef{e:DAE} and \ef{e:sDEQs} with initial conditions which, in general, can be included as one of the parameters.  Then we start with estimated values of $p_k$ and compute \ef{e:dLdp} at the data points $t_i$, then new values of $p_k$ are found from \ef{e:Deltap}, and the process started all over again.  

Petzold has developed a computer code called {\tt DASPK3.0} \cite{LI2000131} to do this for large-scale differential algebraic systems.  Our SEIR fitting problem is not a large scale system, so the code seems to be over kill in our case.

In practice often one fits the log of the population numbers and data rather than the numbers themselves.  Various authors did different things.  
Althaus \cite{Althaus:2014aa} used the {\tt ODE} solver in the R-programming language to solve the SEIR equations and the {\tt optim} optimizer to find the best fit, without explaining what he optimized. 
Tang, \etal\ \cite{Tang:2020aa} used a Markov Chain Monte Carlo (MCMC) method combined with an adaptive Metropolis-Hastings (M-H) algorithm to find the best fit, again without any details.    
The MIT group \cite{Dandekar2020.04.03.20052084} used the local adjoint sensitivity method of Petzold \etal\ \cite{doi:10.1137/S1064827501380630} and Rackauckas \etal\ \cite{rackauckas2020universal} for the machine learning algorithm and the {\tt ADAM} optimizer in the {\tt JULIA} computation environment.  Smirnova \etal\ \cite{Smirnova:2019aa} took a different tack and converted the set of ODE's to a linear Volterra equation which was then solved iteratively for $\beta(t)$.

%
%
\subsubsection{\label{sss:fitexample}Example}

\begin{figure}[t]
  \centering
  \subfigure[\ $I(t)$ (in red) and $S(t)$ (in blue)]
     {\label{f:Fit-Is}
        \includegraphics[width=0.45\columnwidth]{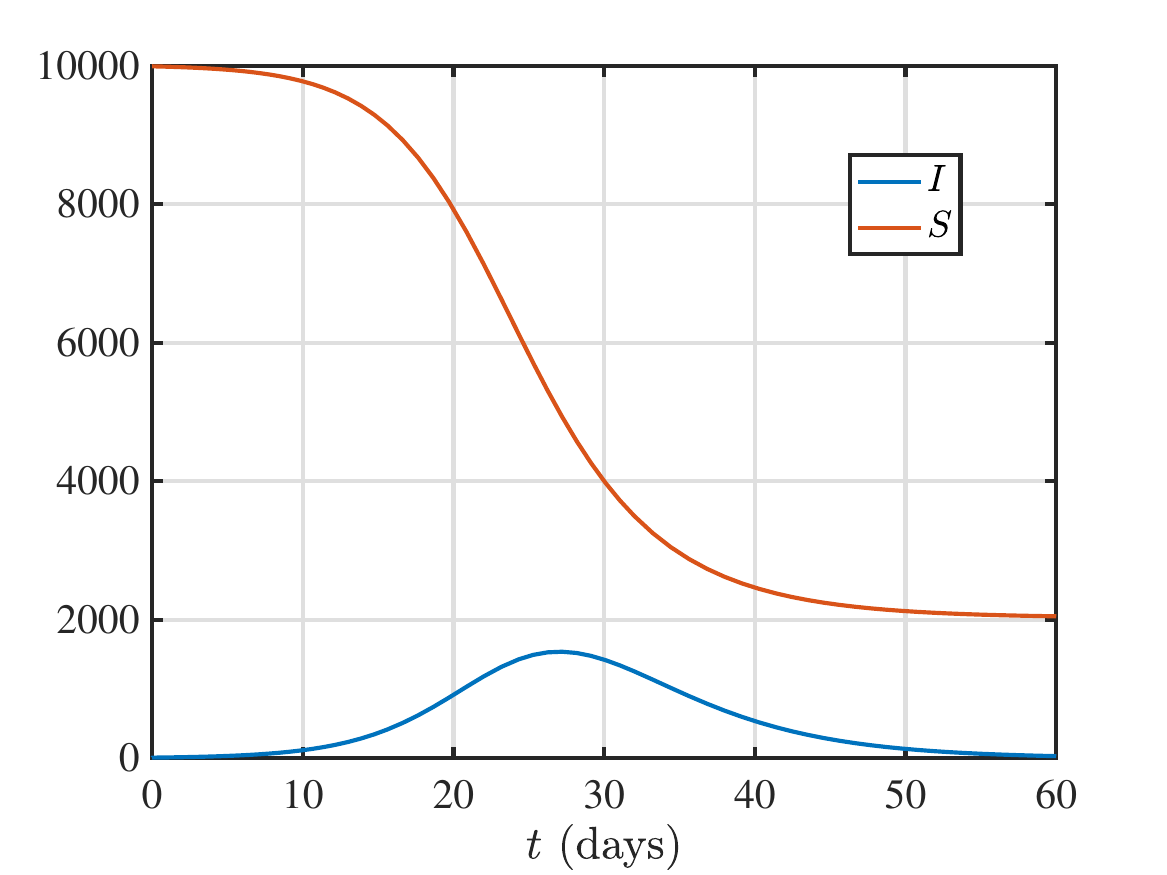} }
  \subfigure[\ $s_{ij}(t)$ ]
     {\label{f:Fit-svalues}
        \includegraphics[width=0.45\columnwidth]{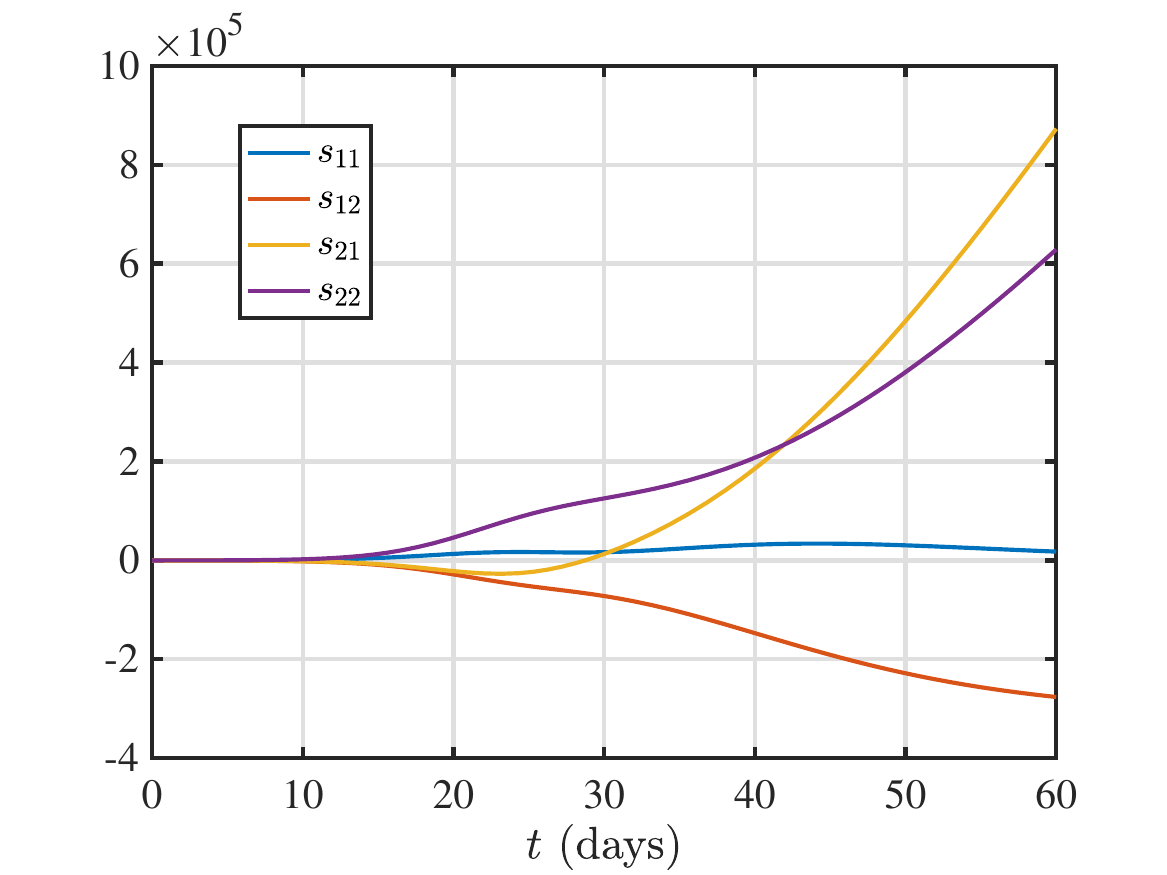} }
  \caption{\label{f:SIRfit} Solutions of Eqs.~\ef{e:yeqs} and 
  \ef{e:dotseqs} with initial conditions \ef{e:fitICs}, for the
  values of $p_1 = \beta = 1/2$ and $p_2 = \gamma = 1/4$.}
\end{figure}

As an example, let us take the SIR model with two model populations $y_1(t) = I(t)$ and $y_2(t) = S(t)$, and with two model (unknown) parameters, $p_1 = \beta$ and $p_2 = \gamma$, satisfying the differential equations,
\begin{subequations}\label{e:yeqs}
\begin{align}
   \dot{y}_1 &= f_1(y,p) = p_1 \, y_1 \, y_2 / N - p_2 \, y_1 \>,
   \label{e:fitfeqs-a} \\
   \dot{y}_2 &= f_2(y,p) = - p_1 \, y_1 \, y_2 / N \>,
   \label{e:fitfeqs-a}
\end{align}
\end{subequations}
with initial conditions, $y_1(0) = I_0$ and $y_2(0) = N - I_0$.  Here we consider $I_0$ and $N$ to be fixed.
So then
\begin{equation}\label{e:dfdy}
   \pdv{f_i}{y_j}
   =
   \begin{pmatrix}
      p_1 y_2 / N - p_2 \>, & p_1 y_1 / N \\
      - p_1 y_2 / N \>, & - p_1 y_1 / N 
   \end{pmatrix} \>,
\end{equation}
and 
\begin{equation}\label{e:dfdp}
   \pdv{f_i}{p_j}
   =
   \begin{pmatrix}
      y_1 \, y_2 / N \>, & - y_1 \\
      - y_1 \, y_2 / N \>, & 0
   \end{pmatrix} \>.
\end{equation}
So then \ef{e:sDEQs} becomes:
\begin{equation}\label{e:dsdt}
   \begin{pmatrix}
     \dot{s}_{11} & \dot{s}_{12} \\
     \dot{s}_{21} & \dot{s}_{22}
   \end{pmatrix}
   =
   \begin{pmatrix}
      p_1 y_2 / N - p_2 \>, & p_1 y_1 / N \\
      - p_1 y_2 / N \>, & - p_1 y_1 / N 
   \end{pmatrix}
   \begin{pmatrix}
     s_{11} & s_{12} \\
     s_{21} & s_{22}
   \end{pmatrix}
   +
   \begin{pmatrix}
      y_1 \, y_2 / N \>, & - y_1 \\
      - y_1 \, y_2 / N \>, & 0
   \end{pmatrix} \>,
\end{equation}
which yields the four equations:
\begin{subequations}\label{e:dotseqs}
\begin{align}
   \dot{s}_{11}
   &=
   [\, p_1 y_2 / N - p_2 \,] \, s_{11}
   +
   [\, p_1 y_1 / N \,] \, s_{21}
   +
   y_1 \, y_2 / N \>,
   \label{e:dots11} \\
   \dot{s}_{21}
   &=
   -
   [\, p_1 y_2 / N \,] \, s_{11}
   -
   [\, p_1 y_1 / N \,] \, s_{21}
   -
   y_1 \, y_2 / N \>,
   \label{e:dots21} \\
   \dot{s}_{12}
   &=
   [\, p_1 y_2 / N - p_2 \,] \, s_{12}
   +
   [\, p_1 y_1 / N \,] \, s_{22}
   -
   y_1 \>,
   \label{e:dots12} \\
   \dot{s}_{22}
   &=
   -
   [\, p_1 y_2 / N \,] \, s_{12}
   -
   [\, p_1 y_1 / N \,] \, s_{22} \>,
   \label{e:dots22}
\end{align}
\end{subequations}
We need to solve the six equations \ef{e:yeqs} and \ef{e:dotseqs}, with the initial conditions,
\begin{equation}\label{e:fitICs}
   y_1(0) = 5
   \qc
   y_2(0) = N - 5
   \qc
   s_{ij}(0) = 0 \>.
\end{equation}
Note that $y_2(0) \gg y_1(0)$.  Solutions of these equations are shown in Fig.~\ref{f:SIRfit}.

%
%
%
%

%
%
\subsubsection{\label{sss:SIREbola}SIR model and the 2014 Ebola epidemic in West Africa}

We try here to fit the simpler SIR model to the 2014 Ebola epidemic in West Africa.  The strict SIR model we use has two parameters $\beta$ and $\gamma$ and is given by the equations,
\begin{subequations}\label{e:SIREbola}
\begin{align}
   \dv{S}{t}
   &=
   - \beta \, S(t) I(t) / N \>,
   \label{e:SIR-a} \\
   \dv{I}{t}
   &=
   \beta \, S(t) I(t) / N - \gamma I(t) \>,
   \label{e:SIR-b} 
\end{align}
\end{subequations}
with initial conditions, $I(0) = I_0$ and $S(0) = N - I_0$.  The total population number $N$ is fixed.  The total number of infected individuals at time $t$ is then given by 
\begin{equation}\label{e:SIR-TotalI}
   C(t)
   =
   N - S(t)
   \qc
   \dv{C}{t} = \beta \, S(t) I(t) / N \>.
\end{equation}
A certain fraction $f$ of these die.

%
%
\subsection{\label{ss:Ebola}Application to the 2014 Ebola epidemic in West Africa}

As an example of application of the SEIR model to a real epidemic, we show an application of this model to the Ebola epidemic in West Africa in 2014.  The epidemic infected about 30,000 people and caused 11,000 deaths in Guinea, Sierra Leone, and Liberia.  

%
%
\begin{figure}[t]
  \centering
  \subfigure[\ Ebola SEIR model]
  { \label{f:EbolaSEIR}
    \includegraphics[width=0.45\columnwidth]{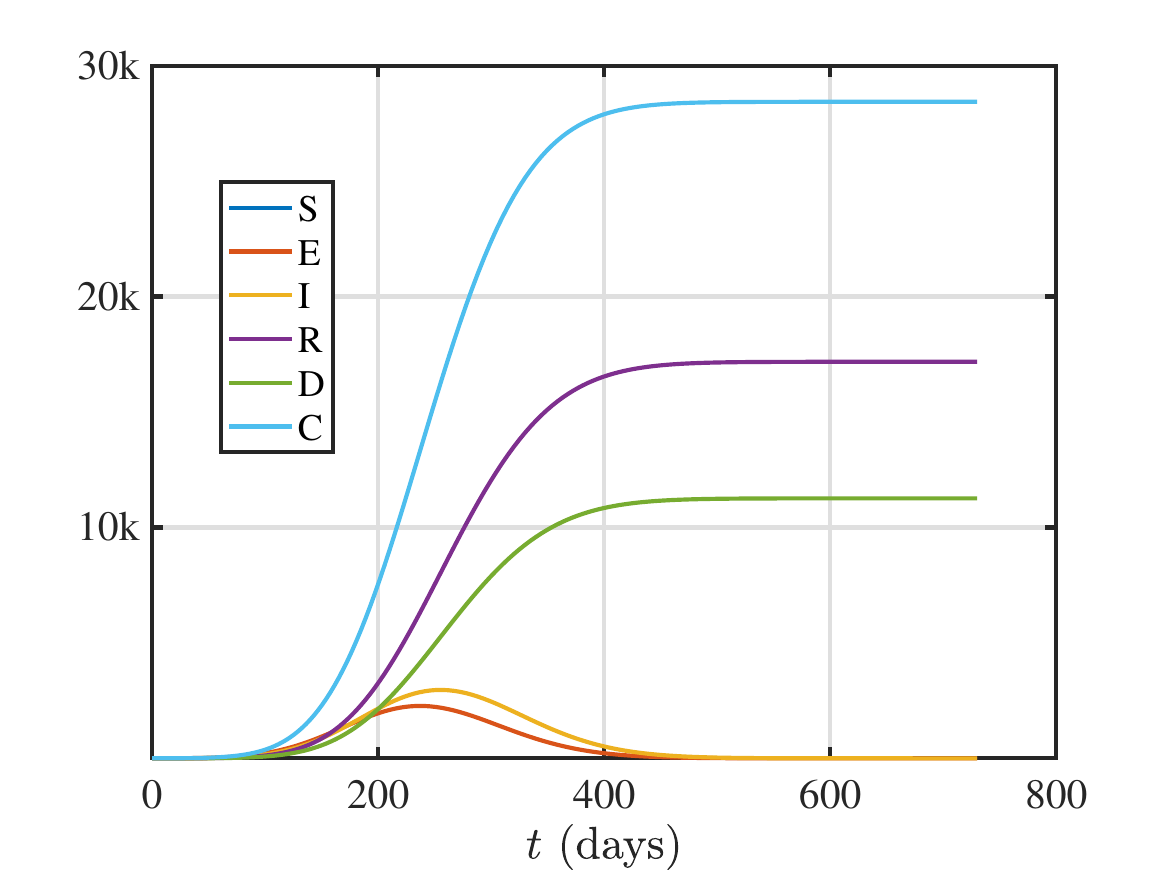} }
  \subfigure[\ New cases and deaths/day]
  { \label{f:EbolaSEIRdetail}
    \includegraphics[width=0.45\columnwidth]{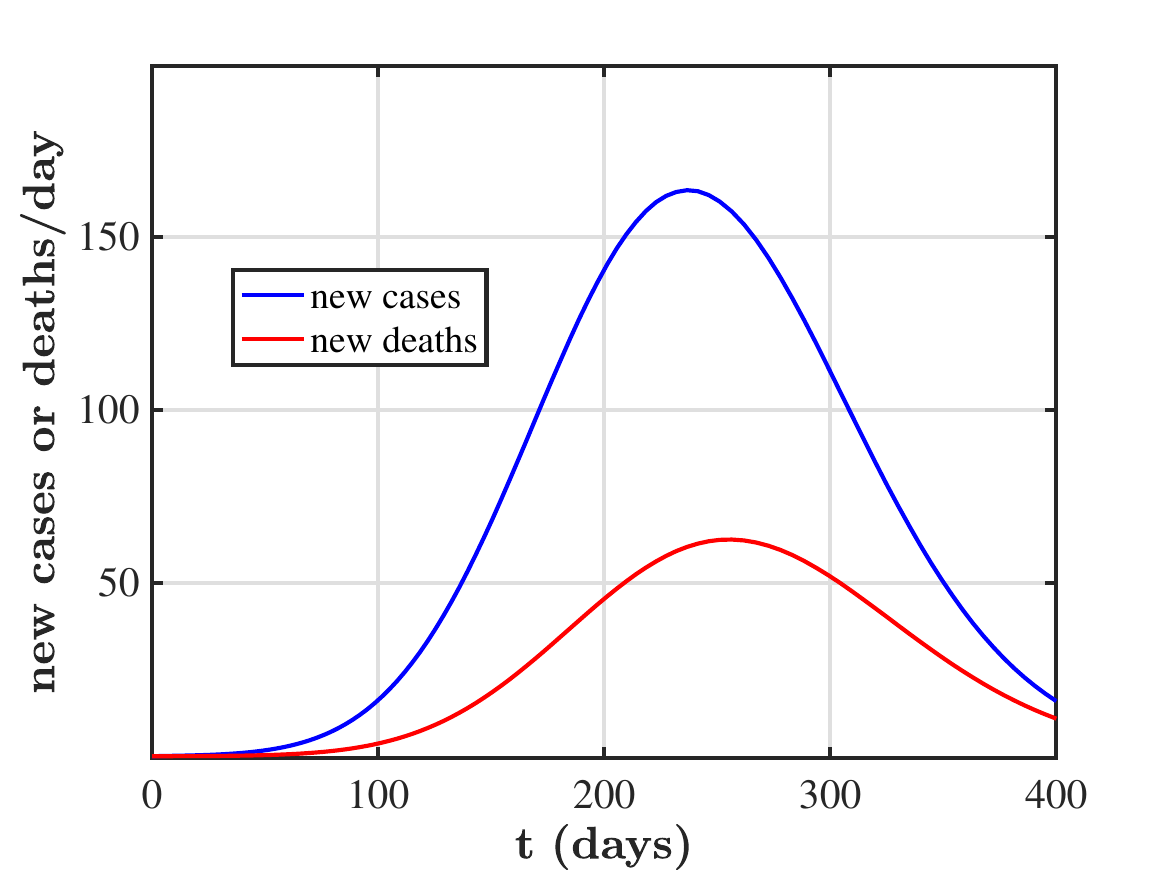} }
  \subfigure[\ Model fit to cases and deaths data]
  { \label{f:EbolaCases}
    \includegraphics[width=0.45\columnwidth]{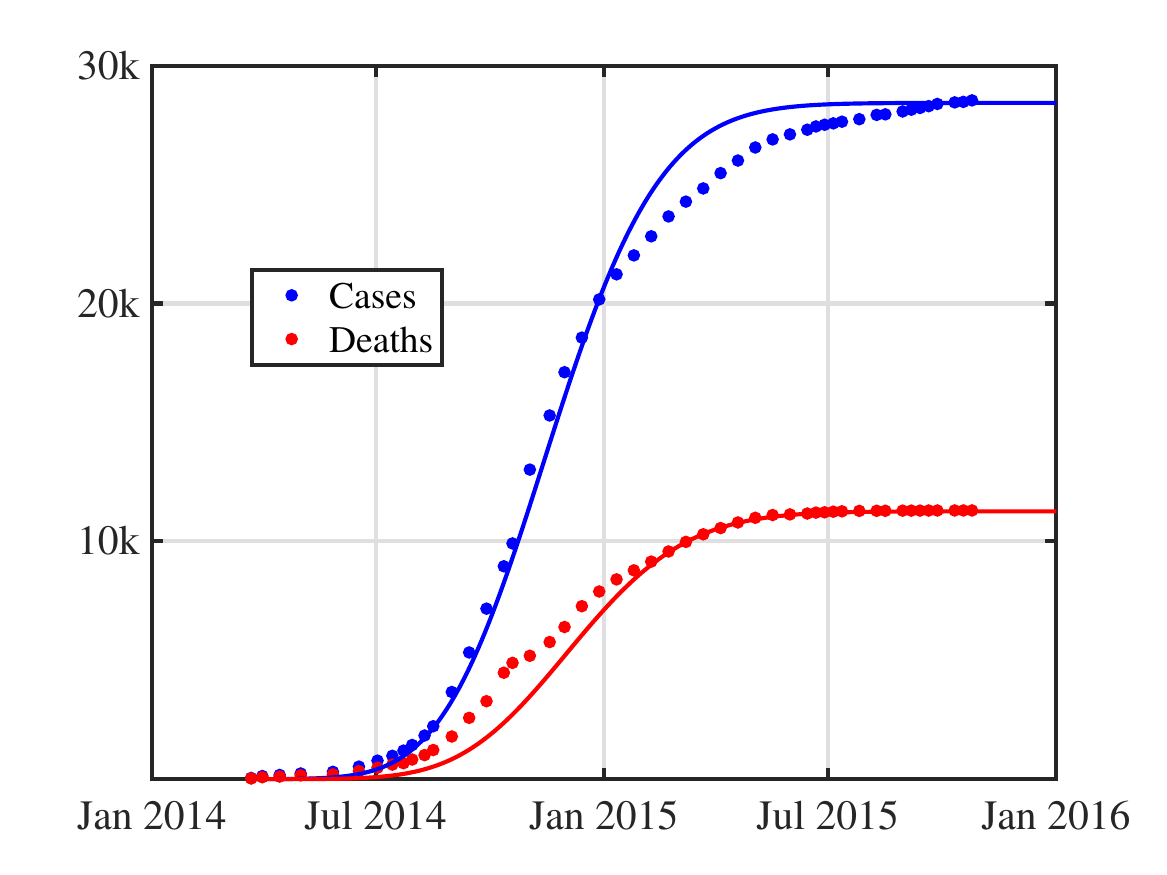} }
  \subfigure[\ $R_{\text{eff}}$ for Ebola epidemic]
  { \label{f:ReffEbola}
    \includegraphics[width=0.45\columnwidth]{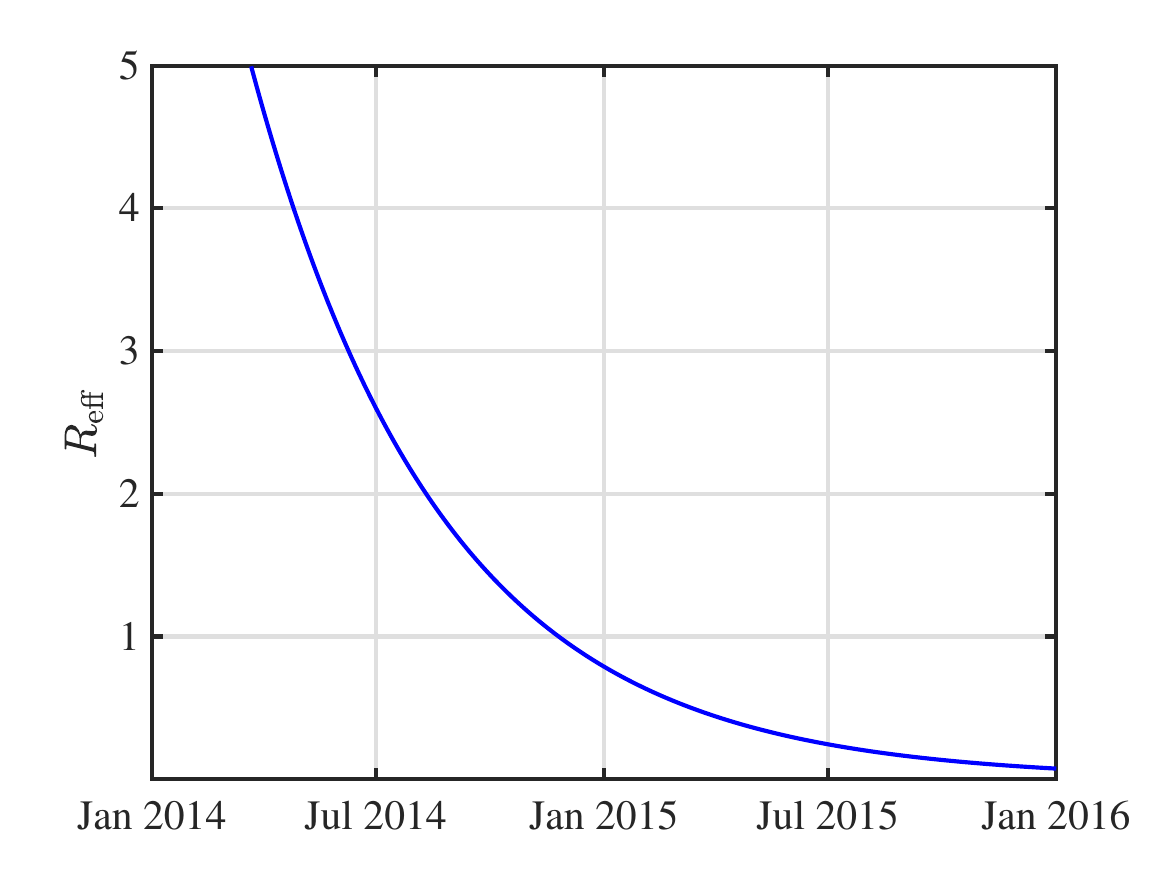} }
  \caption{\label{f:Ebola} The Ebola epidemic in West Africa,
  2014--2016.}
\end{figure}
%
%

To study this case, we modified Eqs.~\ref{e:SEIRbasic} by allowing $\beta$ to be a function of $t$ so as to account for control measures after the epidemic started.  
We used a function $\beta(t)$ proposed by Althais \cite{Althaus:2014aa}, where the infected rate is a constant $\beta_0$ until a time $t_0$ when control measures are taken to reduce the rate of infections,
\begin{equation}\label{e:betaAlthais}
   \beta(t)
   =
   \begin{cases}
      \beta_0 & \text{for $t < t_0$,} \\
      \beta_0 \exp{ - \alpha \, (t - t_0) }
      & \text{for $t \ge t_0$.} 
   \end{cases}
\end{equation}
We started with an initial population of a million susceptible individuals ($S(0) = 10^6$) and with one infected person ($I(0) = 1$) in the population at $t=0$.  For the results shown in Fig.~\ref{f:Ebola}, we used the parameters:
\begin{equation}\label{e:EbolaParms}
   \sigma = 0.0720
   \qc
   \gamma = 0.0533
   \qc
   f = 0.396
   \qc
   \beta_0 = 0.266
   \qc
   \alpha = 0.00648
   \qc
   t_0 = 1 \>.
\end{equation}
The incubation half-life is $\ln{2}/\sigma = 6.72$ days and the infectious half-life is $\ln{2}/\gamma = 13.0$ days.  Figs.~\ref{f:EbolaSEIR} and \ref{f:EbolaSEIRdetail} show the results for $E(t)$, $I(t)$, $R(t)$, $C(t)$, and $D(t)$.  The results for $S(t)$ are off the scale of plots.  The initial transmission half-life is $\ln{2}/\beta_0 = 2.61$ days, but increases rapidly with social distancing.  Fits to WHO data, summarized in Wikipedia\footnote{\url{https://en.wikipedia.org/wiki/West_African_Ebola_virus_epidemic_timeline_of_reported_cases_and_deaths}} for the three countries, is shown in Fig.~\ref{f:EbolaCases}.  
 
The overall Ebola fatality rate ($f$) in West Africa was a shockingly 39.6\% of the cases, much worse than COVID-19.  The control measures response time (social control) was $\ln{2}/\alpha = 107$ days, a poor response to the epidemic.  Adjustment of six parameters are required to do the fit, which was done ``by hand'' since we were not interested in an exact fit.  Each countries statistics and responses to the epidemic were somewhat different, but we used the total results for all three countries.  It is interesting to compare these results, obtained well past the endemic, to those of Althais \cite{Althaus:2014aa} done up to August, 2014, with incomplete data.  It illustrates that it is very difficult to expect much predictive power with an SEIR model until the epidemic has begun to slow down --- lessons one should keep in mind when considering the results of current model predictions in the case of COVID-19.
   
A favored statistical measure is the reproduction number $R$ for the epidemic.  For the Ebola case, the basic reproduction number $R_0 = \beta_0/\gamma = 5$ is well above unity.  However as time goes on with control measures taking place, the effective reproduction number becomes $R_{\text{eff}}(t) = \beta(t) S(t)/ ( \gamma N(t) ) \approx \beta(t)/\gamma$.  This number is plotted in Fig.~\ref{f:ReffEbola} showing the rapid decrease after control measures were taken.  However it was not until about July 22, 2015 that $R_{\text{eff}}(t)$ dropped to below unity.  So it took a long time before the epidemic was brought under control, but it \emph{was} finally brought under control.

%
%
\subsection{\label{ss:USCOVID}Application to the US COVID-19 pandemic}

Application of the SEIR model to the COVID-19 epidemic in the US is difficult since the number of cases are still increasing, although the number of new cases are plateauing or even decreasing.  

When this paper was written (March, 2021), the US was still in the middle of the epidemic and the number of cases were still increasing.  So at that time, application of the SEIR model to the COVID-19 epidemic in the US was difficult.  Regional differences did develop which the simple SEIR model cannot emulate.  However the methods used here could be incorporated into models which better capture the current situation.  

%
%

%
%
\begin{acknowledgments}
JFD would like to thank the Santa Fe Institute for hospitality where some of this work was done.  
\end{acknowledgments}

%
%
\appendix

%
%
\section{\label{s:linearSEIR}Solution of the linearized SEIR equations}

In this appendix, we find general solutions of the linearized SEIR equations \ef{e:SEIRII}.  
Eqs.~\ef{e:SEIRII-b} and \ef{e:SEIRII-c} are of the form,
\begin{equation}\label{e:EImateq}
   \dv{t}
   \begin{pmatrix}
      E(t) \\ I(t)
   \end{pmatrix}
   =
   M \, 
   \begin{pmatrix}
      E(t) \\ I(t)
   \end{pmatrix}
   \qc
   M
   =
   \begin{pmatrix}
      -\sigma & \beta \\
      \sigma & - \gamma
   \end{pmatrix} \>,
\end{equation}
which gives
\begin{equation}\label{e:Lambda}
   \dv[2]{t}
   \begin{pmatrix}
      E(t) \\ I(t)
   \end{pmatrix}
   =
   \Lambda \, 
   \begin{pmatrix}
      E(t) \\ I(t)
   \end{pmatrix}
   \qc
   \Lambda
   =
   M^2
   =
   \begin{pmatrix}
      \beta \sigma + \sigma^2 & -\beta \gamma - \beta \sigma \\
      - \gamma \sigma - \sigma^2 & \gamma^2 + \beta \sigma
   \end{pmatrix} \>.
\end{equation}
Eigenvalues of $\Lambda$ are
\begin{subequations}\label{e:LambdaEigen}
\begin{align}
   \Lambda_{\pm}
   &=
   A \pm B
   \qc
   \lambda_{\pm} = \sqrt{\Lambda_{\pm}} \>,
   \label{e:LambdaEigen-a} \\
   A
   &=
   \frac{1}{2} \, (\, \gamma^2 + 2 \beta \, \sigma + \sigma^2 \,) \>,
   \label{e:LambdaEigen-b} \\
   B
   &=
   \frac{1}{2} \, ( \gamma + \sigma ) \, r
   \qc
   r
   =
   \sqrt{(\gamma - \sigma)^2 + 4 \,\beta \,\sigma} \>,
\end{align}
\end{subequations}
with eigenvectors:
\begin{equation}\label{e:LambdaVectors}
   \chi_{+}
   =
   \begin{pmatrix}
      [\, 
         \gamma - \sigma
         - 
         r \,
      ]/(2 \sigma) \\
      1
   \end{pmatrix}
   \qc
   \chi_{-}
   =
   \begin{pmatrix}
      [\, 
         \gamma - \sigma
         + 
         r \,
      ]/(2 \sigma) \\
      1
    \end{pmatrix} \>.
\end{equation}
The general solution of \ef{e:Lambda} is then
\begin{equation}\label{e:GenSol}
   \begin{pmatrix}
      E(t) \\ I(t)
   \end{pmatrix}
   =
   \chi_{+} \, [\, C_{1} \cosh(\lambda_{+} \, t) + C_{2} \sinh(\lambda_{+} \, t) \, ]
   +
   \chi_{-} \, [\, D_{1} \cosh(\lambda_{-} \, t) + D_{2} \sinh(\lambda_{-} \, t) \, ] \>,
\end{equation}
so that at $t=0$,
\begin{equation}\label{e:BC1}
   \begin{pmatrix}
      E_0 \\ I_0
   \end{pmatrix}
   =
   \chi_{+} \, C_{1}
   +
   \chi_{-} \, D_{1} \>.
\end{equation}
Also from \ef{e:EImateq} at $t=0$,
\begin{equation}\label{e:BC2}
   \begin{pmatrix}
      \dot{E}_0 \\ \dot{I}_0
   \end{pmatrix} 
   =
   M 
   \begin{pmatrix}
      E_0 \\ I_0
   \end{pmatrix}   
   =
   \begin{pmatrix}
      - \sigma \, E_0 + \beta \, I_0 \\
      \sigma \, E_0 - \gamma \, I_0
   \end{pmatrix}
   =
   \chi_{+} \, \lambda_{+} C_{2}
   +
   \chi_{-} \, \lambda_{-} D_{2} \>.   
\end{equation}
The eigenvectors $\chi_{\pm}$ are not orthogonal, however we can construct dual vectors,
\begin{equation}\label{e:duals}
   \psi_{+}
   =
   \begin{pmatrix}
      1 \\
      - [\, \gamma - \sigma - r \,]/(2 \sigma)
   \end{pmatrix} 
   \qc
   \psi_{-}
   =
   \begin{pmatrix}
      1 \\
      - [\, \gamma - \sigma + r \,]/(2 \sigma)
    \end{pmatrix} \>,
\end{equation}
which obey:
\begin{subequations}\label{e:dualprops}
\begin{gather}
   \psi_{+}^T \chi_{+} = 0
   \qc
   \psi_{+}^T \chi_{-} = 
   + r / \sigma \>,
   \\
   \psi_{-}^T \chi_{-} = 0
   \qc
   \psi_{-}^T \chi_{+} = 
   - r/ \sigma \>,
\end{gather}
\end{subequations}
and can be used to invert \ef{e:BC1} and \ef{e:BC2}.  This gives
\begin{subequations}\label{e:invert}
\begin{align}
   C_1
   &=
   - \frac{\sigma}{r} \,
   \psi_{-}^T 
   \begin{pmatrix}
      E_0 \\ I_0
   \end{pmatrix}
   =
   -
   \frac{\sigma}{r} E_0
   +
   \frac{1}{2} \,
   \Bigl \{\,
      1
      +
      \frac{\gamma - \sigma}{r} \,
   \Bigr \} \, I_0 \>,
   \label{e:invert-a} \\
   D_1
   &=
   + \frac{\sigma}{r} \,
   \psi_{+}^T 
   \begin{pmatrix}
      E_0 \\ I_0
   \end{pmatrix}
   =
   +
   \frac{\sigma}{r} E_0
   +
   \frac{1}{2} \, 
   \Bigl \{\,
      1
      -
      \frac{\gamma - \sigma}{r} \,
   \Bigr \} \, I_0 \>.
   \label{e:invert-b} \\
   C_2
   &=
   - \frac{\sigma}{r \, \lambda_{+} } \,
   \psi_{-}^T 
   \begin{pmatrix}
      \dot{E}_0 \\ \dot{I}_0
   \end{pmatrix}
   =
   -
   \frac{\sigma}{r \, \lambda_{+}} \dot{E}_0
   +
   \frac{1}{2 \, \lambda_{+}} \,
   \Bigl \{\,
      1
      +
      \frac{\gamma - \sigma}{r} \,
   \Bigr \} \, \dot{I}_0 \>,
   \label{e:invert-c} \\
   D_2
   &=
   + \frac{\sigma}{r \, \lambda_{-}} \,
   \psi_{+}^T 
   \begin{pmatrix}
      \dot{E}_0 \\ \dot{I}_0
   \end{pmatrix}
   =
   +
   \frac{\sigma}{r\, \lambda_{-}} \dot{E}_0
   +
   \frac{1}{2 \, \lambda_{-}} \,
   \Bigl \{\,
      1
      -
      \frac{\gamma - \sigma}{r} \,
   \Bigr \} \, \dot{I}_0 \>.
   \label{e:invert-d}
\end{align}
\end{subequations}
Solutions are then given by \ef{e:GenSol}.
Populations of $F(t)$, $C(t)$, $R(t)$ and $D(t)$ can be found from \ef{e:GenSol} by integration.  

%
%
\section{\label{s:Langevin}Derivation of the Langevin equations}

Derivation of the Langevin equations from a microscopic model follow a well-known path, described in the flow diagram below:
\bigskip
\begin{center}
\begin{tikzpicture}
  [node distance=.8cm, start chain=going below,]
     \node[punktchain, join] (intro) {Microscopic model};
     \node[punktchain, join] (probf) {Master equation};
     \node[punktchain, join] (investeringer) {Schrodinger equation};
     \node[punktchain, join] (perfekt) {Continuum variables};
     \node[punktchain, join, ] (emperi) {Path integral};
     \node[punktchain, join, ] (emperi) {Langevin equations};
\end{tikzpicture}
\end{center}
\bigskip
We show how this is done in this appendix, starting with the master equation.  

%
%
\subsection{\label{ss:master}The master equation} 

The master equation formalism assumes that the space in which reactions take place can be divided into a $d$-dimensional hyper-cubic lattice of cells and that each cell can be treated as a coherent entity.  Reactions occur only if the objects involved are in the same cell.  They are assumed to react independently.  If the underlying processes are Markovian, they can be described by a probability distribution function $P(\, \vb{N},t \,)$, where $\vb{N} = \{\, N_{a}(i), N_{b}(j), \dotsc \,\}$ is a vector describing the number of objects $N_{\alpha}(i) $ of species $\alpha = (a,b,\dotsc)$ at a site $i$ at time $t$.  For chemical reactions, the master equation given by the equation \cite{r:Roussel:2007fk}
\begin{equation}\label{ME.e:1}
   \frac{\partial P(\vb{N},t)}{\partial t}
   =
   \sum_{r} \,
   \bigl [ \,
      a_r(\vb{N} - \vb{S}_r) \, P(\vb{N}-\vb{S}_r,t)
      -
      a_r(\vb{N}) \, P(\vb{N},t) \,
   \bigr ] \>,
\end{equation}
where $\vb{S}_r = \{\, S_S,S_E,S_I,S_R \,\}$ is the stoichiometric vector of reaction $r$ and $a_r(\vb{N})$ is the propensity function (reaction rate), determined by the number and kind of objects present before the reaction.  From \eqref{ME.e:1}, we find immediately that $\sum_{\vb{N}} P(\vb{N},t) = 1$ is conserved.

%
%
\subsection{\label{ss:manybody}Many-body formulation}

The master equation suggests introducing an occupation number algebra with annihilation and creation operators\footnote{We designate operators by a circumflex.} $\hat{a}^{\phantom\dag}_{\alpha}(i)$ and $\hat{a}^{\dag}_{\alpha}(i)$ for each site and for each species, where the index $i$ labels the site and the index $\alpha$ the species.  The operators obey the commutation relations
\begin{equation}\label{GS.e:3}
   \comm*{ \hat{a}^{\phantom\dag}_{\alpha}(i) }
        { \hat{a}^{\dag}_{\beta}(j) } 
   = 
   \delta_{\alpha,\beta} \, \delta_{i,j} \>,
\end{equation}
with number operators of the form 
\begin{equation}\label{GS.e:4}
   \hat{N}^{\phantom\dag}_{\alpha}(i)
   =
   \hat{a}^{\dag}_{\alpha}(i) \,\hat{a}^{\phantom\dag}_{\alpha}(i)
\end{equation}
for each site and each species, which have integer eigenvalues,
\begin{equation}\label{GS.e:5}
   \hat{N}^{\phantom\dag}_{\alpha}(i) \,
   \ket{N^{\phantom\dag}_{\alpha}(i) }
   =
   N^{\phantom\dag}_{\alpha}(i) \,
   \ket{ N^{\phantom\dag}_{\alpha}(i) } \>,
\end{equation}
with $N^{\phantom\dag}_{\alpha}(i) = 0,1,2,\dotsc$.  The vacuum state is defined by $\hat{a}^{\phantom\dag}_{\alpha}(i) \, \ket{\bzero} = \bra{\bzero} \, \hat{a}^{\dag}_{\alpha}(i) = 0$ for 
all $i$ and $\alpha$.
In this occupation number space, we define a state vector $\ket{\Psi(t)}$ by:
\begin{equation}\label{ME.e:2}
   \ket{\Psi(t)}
   =
   \prod_{\alpha,i} \sum_{N_{\alpha}(i) = 0}^{\infty} 
   \! P(\vb{N})
   \bigl [ \,
      \hat{a}^{\dag}_{\alpha}(i) \,
   \bigr ]^{N^{\phantom\dag}_{\alpha}(i)} \,
   \ket{\vb{0}} \>.
\end{equation}
The master equation can then be used to find a rate equation for the state vector $\ket{\Psi(t)}$, which is of a ``\Schrodinger'' form in imaginary time,
\begin{equation}\label{ME.e:3}
   \pdv{\ket{\Psi(t)}}{t}
   =
   - 
   H[\, \hat{\vb{a}}, \hat{\vb{a}}^{\dag} \, ] \,
   \ket{\Psi(t)} \>.
\end{equation}
We refer to the rate operator $H[\, \hat{\vb{a}}, \hat{\vb{a}}^{\dag} \, ]$ as the ``Hamiltonian'' operator, in analogy to quantum mechanics, although this operator is not hermitian.  Note the minus sign in the definition of $H$.
Contributions to the master equation come from the hopping of species between nearest neighbor sites, and the reactions between the species.  For our case, the probability of finding values $\vb{N} = \{\, S, E, I, R \,\}$ of the populations is given by a function  $P( \vb{N} )$.  The reaction parameters for the SEIR model are given in Table~\ref{t:Reactions}.  We work out the contributions of each reaction to the master equation and Hamiltonian in the following.

%
%
\begin{table}[t]
\centering
\caption{\label{t:Reactions} Reaction parameters for the SIR model}
\begin{ruledtabular}
\begin{tabular}{c @{\qquad} c @{\qquad} c @{\qquad} c @{\qquad} c @{\qquad} c}
   $r$  & reaction & $a(\vb{N})$ & rate & $\vb{S} = (S,I)$ \\
   \hline
   1 & $S + I \xrightarrow{k_1} 2 \, I$ & $k_1 \, S \, I$ & 
        $\lambda= k_1 L^{d} = \beta/\rho_0$ & $(-1,+1)$  \\
   2 & $I \xrightarrow{k_2} 0$ & $k_2 \, I$ & $\mu = k_2 L^d$ & $(0,-1)$  \\
   3 & $S \xrightarrow{k_3} 0$ & $k_3 \, S$ & $\nu = k_3 L^d$ & $(-1,0)$ \\
   4 & $0 \xrightarrow{k_4} S$ & $k_4$ & $f = k_4$ & $(1,0)$  \\
   6 & $U_i \xrightarrow{d} U_{i \pm 1}$ & $d\, U_i$ & $D = 2 \, d / h^2$ & $U_i \pm 1$ \\
\end{tabular}
\end{ruledtabular}
\end{table}
%
%

\begin{enumerate}
%
%
\item For the reaction $S + I \xrightarrow{k_1} 2 I$, the stoichiometric vector is $\vb{S} = (-1,+1)$ and the reaction rate is $a = k_1 \, S \, I$.  So this event contributes a factor of
\begin{equation}\label{reaction-k1}
   k_1 \,
   \bigl \{\,
      (S + 1) \, (I - 1) \, P(S+1,I-1)
      -
      S \, I \, P(S,I) \,
   \bigr \} \>,
\end{equation}
to the master equation at each site $i$, and a factor
\begin{equation}\label{H-reaction-k1}
   k_1 \sum_{i}\,
   [\, \hat{a}_S^{\dag}(i) - \hat{a}_I^{\dag}(i) \,] \,
   \hat{a}_I^{\dag}(i) \, \hat{a}_S(i) \, \hat{a}_I(i)
\end{equation}
to the Hamiltonian.
%
%
\item For the reaction $I \xrightarrow{k_2} 0$, the stoichiometric vector is $\vb{S} = (0,-1)$, and the reaction rate is $a = k_2 \, I$.  So this event contributes a factor of
\begin{equation}\label{reaction-k3}
   k_2 \,
   \bigl \{\,
      (I + 1) \, P(S,I+1)
      -
      I \, P(S,I) \,
   \bigr \} \>,
\end{equation}
to the master equation at each site $i$, and a factor
\begin{equation}\label{H-reaction-k3}
   k_2 \sum_{i} \, 
   [\, \hat{a}_I^{\dag}(i) - 1 \,] \, \hat{a}_I(i)
\end{equation}
to the Hamiltonian.
%
%
\item For the reaction $S \xrightarrow{k_3} 0$, the stoichiometric vector is $\vb{S} = (-1,0)$, and the reaction rate is $a = k_3 \, S$.  So this event contributes a factor of
\begin{equation}\label{reaction-k3}
   k_3 \,
   \bigl \{\,
      (S + 1) \, P(S+1,I)
      -
      S \, P(S,I) \,
   \bigr \} \>,
\end{equation}
to the master equation at each site $i$, and a factor
\begin{equation}\label{H-reaction-k3}
   k_3 \sum_{i} \, 
   [\, \hat{a}_S^{\dag}(i) - 1 \,] \, \hat{a}_S(i)
\end{equation}
to the Hamiltonian.
%
%
\item For the reaction $0 \xrightarrow{k_4} S$, the stoichiometric vector is $\vb{S} = (1,0)$, and the reaction rate is $a = k_4$.  So this event contributes a factor of
\begin{equation}\label{reaction-k4}
   k_4 \,
   \bigl \{\,
      P(S-1,I)
      -
      P(S,I) \,
   \bigr \} \>,
\end{equation}
to the master equation at each site $i$, and a factor
\begin{equation}\label{H-reaction-k4}
    k_4 \sum_{i} \, [\, 1 - \hat{a}_S^{\dag}(i)\,]
\end{equation}
to the Hamiltonian.
%
%
\item Finally, the diffusion process is generated by hopping of like species between nearest neighbor sites $1$ and $2$ at a rate $d_{\alpha}$.  If $n_1$ is the number of particles of species $\alpha$ at site $1$ and $n_2$ the number of particles of species $\alpha$ at site $2$, this event contributes a factor
\begin{equation*}
   d_{\alpha} \, 
   \bigl \{ \,
      (n_1 + 1) \, P( n_1 + 1, n_2 - 1, t ) - n_1 \, P(n_1,n_2,t) \,
   \bigr \}
\end{equation*}
to the master equation for each species, and a corresponding factor 
\begin{equation*}
   d_{\alpha} \, 
   ( \, 
      \hat{a}^{\dag}_{\alpha}(2) \, \hat{a}_{\alpha}(1) 
      - 
      \hat{a}^{\dag}_{\alpha}(1) \, \hat{a}_{\alpha}(1) \, 
   )
\end{equation*}
to the Hamiltonian operator.  Similarly, hopping of atoms from nearest neighbor sites $2$ to $1$ gives a factor
\begin{equation*}
   d_{\alpha} \, 
   ( \, 
      \hat{a}^{\dag}_{\alpha}(1) \, \hat{a}_{\alpha}(2) 
      - 
      \hat{a}^{\dag}_{\alpha}(2) \, \hat{a}_{\alpha}(2) \, 
   ) \>.
\end{equation*}
So hopping of species $\alpha$ in both directions contributes a factor
\begin{equation}\label{ME.e:4}
   d_{\alpha} \sum_{\ev{i,j}} \,
   ( \, \hat{a}^{\dag}_{\alpha}(i) - \hat{a}^{\dag}_{\alpha}(j) \, ) \, 
   ( \, \hat{a}_{\alpha}(i) - \hat{a}_{\alpha}(j) \, ) \>, 
\end{equation}
to the Hamiltonian operator for each species.  Here the sum goes over all nearest neighbor sites.  In our model, we allow hopping for the S, E, and I populations.  

\end{enumerate}
%
%

Collecting all terms, the Hamiltonian is given by
\begin{align}
   H[ \hat{\vb{a}}^{\dag},\hat{\vb{a}} ]
   &=
   \sum_{\ev{i,j}} \,
   \bigl \{\,
   d_{I} \,
   ( \, \hat{a}^{\dag}_{I}(i) - \hat{a}^{\dag}_{I}(j) \, ) \, 
   ( \, \hat{a}_{I}(i) - \hat{a}_{I}(j) \, )
   \notag \\[-8pt]
   & \hspace{4em}
   +
   d_{S} \,
   ( \, \hat{a}^{\dag}_{S}(i) - \hat{a}^{\dag}_{S}(j) \, ) \, 
   ( \, \hat{a}_{S}(i) - \hat{a}_{S}(j) \, ) 
   \bigr \}
   \label{HamI} \\[4pt]
   & \hspace{0em}
   +
   \sum_i
   \bigl \{\, 
      k_1 \, [\, \hat{a}_S^{\dag}(i) - \hat{a}_I^{\dag}(i) \,] \,
             \hat{a}_I^{\dag}(i) \, \hat{a}_S(i) \, \hat{a}_I(i)
      +
      k_2 \, [\, \hat{a}_I^{\dag}(i) - 1 \,] \, \hat{a}_I(i)
      \notag \\[-3pt]
      & \hspace{4em}
      +
      k_3 \, [\, \hat{a}_S^{\dag}(i) - 1 \,] \, \hat{a}_S(i)
      +
      k_4 \, [\, 1 - \hat{a}_S^{\dag}(i)\,] \,
   \bigr \} \>.
   \notag
\end{align}
%
%
\subsection{\label{ss:DoiShift}The Doi shift}

A simple trick, discovered by Doi \cite{0305-4470-9-9-008}, provides a probabilistic interpretation of the state vector.  Doi found that if all creation operators are translated by one unit leaving the annihilation operators unchanged, the overlap of the transformed state vector with the vacuum is conserved.  The non-hermitian operator $\hat{D}$ which does this is given by
\begin{equation}\label{DS.e:1}
   \hat{D}
   =
   \expB{ - \sum_{i,\alpha} \hat{a}_{\alpha}(i) } \>,
\end{equation}
so that
\begin{subequations}\label{DS.e:2}
\begin{align}
   \hat{D}^{-1} \,
   \hat{a}_{\alpha}(i) \,
   \hat{D}
   &=
   \hat{a}_{\alpha}(i) \>,
   \label{DS.e:2-a} \\
   \hat{D}^{-1} \,
   \hat{a}^{\dag}_{\alpha}(i) \,
   \hat{D}
   &=
   1 +
   \hat{a}^{\dag}_{\alpha}(i) \>.
   \label{DS.e:2-b} 
\end{align}
\end{subequations}
The Doi-shifted state vector is defined by,
\begin{align}\label{DS.e:2.1}
   \ket*{ \tilde{\Psi}(t) }
   =
   \hat{D}^{-1} \ket*{ \Psi(t) }
   &=
   \prod_{\alpha,i} \sum_{n_{\alpha}(i) = 0}^{\infty} 
   \! P(\vb{n},t) \,
   \hat{D}^{-1} \,
   \bigl [ \, 
      \hat{a}^{\dag}_{\alpha}(i) \, 
   \bigr ]^{n^{\phantom\dag}_{\alpha}(i)} \,
   \hat{D} \, \hat{D}^{-1} \,
   \ket{ \bzero }
   \\
   &=
   \prod_{\alpha,i} \sum_{n_{\alpha}(i) = 0}^{\infty} 
   \! P(\vb{n},t) \,
   \bigl [ \, 
      1 + \hat{a}^{\dag}_{\alpha}(i) \, 
   \bigr ]^{n^{\phantom\dag}_{\alpha}(i)} \,
   \ket{ \bzero } \>,
   \notag
\end{align}
and the overlap of the vacuum and the shifted state vector is given by
\begin{equation}\label{DS.e:3}
   \braket*{ \vb{0} }{ \tilde{\Psi}(t) }
   =
   \prod_{\alpha,i} \sum_{n_{\alpha}(i) = 0}^{\infty} 
   \! P(\vb{n},t) = 1 \>,
\end{equation}
and is conserved.  We can define a Doi-shifted vacuum by $\bra{\tilde{\vb{0}}} = \bra{\vb{0}} \hat{D}^{-1}$ so that $\braket*{ \vb{0} }{ \tilde{\Psi}(t) } = \braket*{ \tilde{\vb{0}} }{ \Psi(t) }$.
The \Schrodinger\ equation \ef{ME.e:3} for the Doi-shifted state vector becomes
\begin{equation}\label{DS.e:13}
   \pdv{\ket*{\tilde{\Psi}(t)}}{t}
   =
   - 
   \tilde{H}[\, \hat{\vb{a}}^{\dag},\hat{\vb{a}} \, ]
   \ket*{\tilde{\Psi}(t)} \>,
\end{equation}
where the shifted Hamiltonian is
\begin{align}
   H[ \hat{\vb{a}}^{\dag},\hat{\vb{a}} ]
   &=
   \sum_{\ev{i,j}} \,
   \bigl \{\,
   d_{I} \,
   ( \, \hat{a}^{\dag}_{I}(i) - \hat{a}^{\dag}_{I}(j) \, ) \, 
   ( \, \hat{a}_{I}(i) - \hat{a}_{I}(j) \, )
   \notag \\[-8pt]
   & \hspace{4em}
   +
   d_{S} \,
   ( \, \hat{a}^{\dag}_{S}(i) - \hat{a}^{\dag}_{S}(j) \, ) \, 
   ( \, \hat{a}_{S}(i) - \hat{a}_{S}(j) \, ) 
   \bigr \}
   \label{HamII} \\[4pt]
   & \hspace{0em}
   +
   \sum_i
   \bigl \{\, 
      k_1 \, [\, \hat{a}_S^{\dag}(i) - \hat{a}_I^{\dag}(i) \,] \,
             (\, \hat{a}_I^{\dag}(i) + 1 \,) \, \hat{a}_S(i) \, \hat{a}_I(i)
      +
      k_2 \, \hat{a}_I^{\dag}(i) \, \hat{a}_I(i)
      \notag \\[-3pt]
      & \hspace{4em}
      +
      k_3 \, \hat{a}_S^{\dag}(i) \, \hat{a}_S(i)
      -
      k_4 \, \hat{a}_S^{\dag}(i) \,
   \bigr \} \>.
   \notag
\end{align}
%
%
\subsection{\label{ss:continuum}The continuum limit}

If the lattice sites are located at $\vb{x}_i = h \, \vb{i}$, where $h$ is the lattice spacing,   
we pass to a continuum limit by setting
\begin{equation}\label{RD.e:A7}
   \sum_i 
   \mapsto
   \int \frac{\dd[d]{x}}{L^d} \>,
   \qquad
   \delta_{i,i'}
   \mapsto
   L^d \, \delta^d(\vb{x} - \vb{x}' ) \>,
\end{equation}
where $L^d$ is the sample volume.  So if we define
\begin{equation}\label{RD.e:A8}
   \hat{a}_{\alpha}(i)
   \mapsto 
   L^d \, \hat{\phi}_{\alpha}(\vb{x}) \>,
   \qquad
   \hat{a}^{\dag}_{\alpha}(i)
   \mapsto
   \hat{\phi}^{\dag}_{\alpha}(\vb{x}) \>,
\end{equation}
then
\begin{equation}\label{RD.e:A9}
   \comm*{ \hat{\phi}_{\alpha}(\vb{x}) }{ \hat{\phi}^{\dag}_{\beta}(\vb{x}') }
   =
   \delta_{\alpha,\beta} \,
   \delta^d(\vb{x} - \vb{x}') \>.
\end{equation}
That is unlike quantum mechanics, $\hat{\phi}_{\alpha}(\vb{x})$ has units of density $1/L^d$, whereas $\hat{\phi}^{\dag}_{\alpha}(\vb{x})$ has no units.  The vacuum is defined by $\hat{\phi}_{\alpha}(\vb{x})\,\ket{\vb{0}} = 0$ and $\bra{\vb{0}}\,\hat{\phi}^{\dag}_{\alpha}(\vb{x}) = 0$.

Setting $\vb*{\phi} = \{\, \phi_S, \phi_E, \phi_I, \phi_R \,\}$, the continuum state vector satisfies the Doi-shifted \Schrodinger-like equation,
\begin{equation}\label{PI.e:SchrodingerII}
   \pdv{\ket*{\tilde{\Psi}(t)}}{t}
   =
   - 
   \tilde{H}[\, \hat{\vb{\vb*{\phi}}}^{\dag},\hat{\vb*{\phi}} \, ]
   \ket*{\tilde{\Psi}(t)} \>.   
\end{equation}
where the Doi-shifted Hamiltonian \eqref{HamII} becomes:
\begin{align}\label{HamIII}
   \tilde{H}[\, \hat{\vb{\vb*{\phi}}}^{\dag},\hat{\vb*{\phi}} \, ]
   &=
   \int \! \dd[d]{x} \,
   \bigl \{\,
   D_I \, [\, \grad{\hat{\phi}^{\dag}_I}(\vb{x}) \,] \vdot 
          [\, \grad{\hat{\phi}_I}(\vb{x}) \,]
   +
   D_S \, [\, \grad{\hat{\phi}^{\dag}_S}(\vb{x}) \,] \vdot 
          [\, \grad{\hat{\phi}_S}(\vb{x}) \,]
   \\
   & \hspace{-4em}
   +
   \lambda \,
   [\, \hat{\phi}_S^{\dag}(\vb{x}) - \hat{\phi}_I^{\dag}(\vb{x}) ] \,
   [\, \hat{\phi}_I^{\dag}(\vb{x}) - 1 \,] \,
   \hat{\phi}_S(\vb{x}) \hat{\phi}_I(\vb{x})
   +
   \mu \, \hat{\phi}_I^{\dag}(\vb{x}) \, \hat{\phi}_I(\vb{x})
   +
   \nu \, \hat{\phi}_S^{\dag}(\vb{x}) \, \hat{\phi}_S(\vb{x})
   -
   f \, \hat{\phi}_S^{\dag}(\vb{x}) \,
   \bigr \} \>.
   \notag
\end{align}
where the constants $(\lambda,\mu,\nu,f)$ are defined in Table~\ref{t:Reactions}.
Integrating by parts, we can rewrite \ef{HamIII} in the form:
\begin{align}\label{HamIII}
   \tilde{H}[\, \hat{\vb{\vb*{\phi}}}^{\dag},\hat{\vb*{\phi}} \, ]
   &=
   -
   \int \! \dd[d]{x} \,
   \bigl \{\,
      \hat{\phi}_I^{\dag}(\vb{x}) \,\,
      \bigl [\,
         \bigl (\, 
            D_I \laplacian 
            -
            \mu \,
         \bigr ) \, \hat{\phi}_I(\vb{x}) \,
         +
         \lambda \, \hat{\phi}_S(\vb{x}) \, \hat{\phi}_I(\vb{x})
      \bigr ] 
      \\
      & \hspace{4.5em}
      \hat{\phi}_S^{\dag}(\vb{x}) \,
      \bigl [\,
         \bigl (\, 
            D_S \laplacian 
            -
            \nu \,
         \bigr ) \, \hat{\phi}_S(\vb{x})
         -
         \lambda \, \hat{\phi}_S(\vb{x}) \, \hat{\phi}_I(\vb{x})
         +
         f \,
      \bigr ]
      \notag \\[7pt]
      & \hspace{4.5em}
      +
      \lambda \,            
      [\, \hat{\phi}_I^{\dag}(\vb{x}) - \hat{\phi}_S^{\dag}(\vb{x}) \,] \,
      \hat{\phi}_I^{\dag}(\vb{x}) \,
      \hat{\phi}_S(\vb{x}) \hat{\phi}_I(\vb{x})
   \bigr \} \>.
   \notag
\end{align}
The last term represents noise.  

%
%
\subsection{\label{ss:pathint}Path integral}

The path integral in a coherent representation for the dynamics is obtained in the usual way.  The Heisenberg transformation bracket can be written as
\begin{equation}\label{e:pathintI}
   \braket*{ \vb*{\phi}_f,t_f }{ \vb*{\phi}_i,t_i }
   =
   \frac{1}{Z(t_f,t_i)} 
   \iint_{\vb*{\phi}(t_i)=\vb*{\phi}_i}^{\vb*{\phi}(t_f)=\vb*{\phi}_f } 
   \frac{ \DD{\vb*{\phi}} \DD{\vb*{\phi}^{\star}} }{ 2\pi i }
   \exp{ - S[\,\vb*{\phi},\vb*{\phi}^{\star} \,] } \>,
\end{equation}
where the action is given by
\begin{align}\label{e:actionI}
   S[\,\vb*{\phi},\vb*{\phi}^{\star} \,]
   &=
   \int \! \dd{x} \,
   \bigl \{\,
      \phi_I^{\star}(x) \,
      \bigl [\,
         \bigl (\,
            \partial_t
            - 
            D_I \laplacian 
            +
            \mu \,
         \bigr ) \, \phi_I(x)
         -
         \lambda \, \phi_S(x) \, \phi_I(x) \,
      \bigr ] 
      \\
      & \hspace{3em}
      +
      \phi_S^{\star}(x) \,
      \bigl [\,
         \bigl (\,
            \partial_t
            - 
            D_S \laplacian 
            +
            \nu \,
         \bigr ) \, \phi_S(x)
         +
         \lambda \, \phi_S(x) \, \phi_I(x)
         -
         f \,
      \bigr ]
      \notag \\[7pt]
      & \hspace{3em}
      -
      \lambda \,            
      [\, \phi_I^{\star}(x) - \phi_S^{\star}(x) \,] \,
      \phi_I^{\star}(x) \, \phi_S(x) \phi_I(x)
   \bigr \} \>.
   \notag
\end{align}
Here we have set $\dd{x} \equiv \dd{t} \dd[d]{x}$.  Recall that for imaginary time, $\phi^{\star}_{\alpha}(x)$ is \emph{not} the adjoint of $\phi_{\alpha}(x)$.  
Equation \ef{e:actionI} is of the form,
\begin{equation}\label{e:actionII}
    S[\,\vb*{\phi},\vb*{\phi}^{\star} \,]
    =
    \iint \dd{x} \dd{x'}
    \bigl \{\,
       \vb*{\phi}^{\star}(x) \vdot \vb{L}[\vb*{\phi}](x,x')
       +
       \frac{1}{2} \, 
       \vb*{\phi}^{\star}(x) \vdot \vb{D}[\vb*{\phi}](x,x') \vdot 
       \vb*{\phi}^{\star}(x) \,
    \bigr \} \>,
\end{equation}
where $\vb{L}[\vb*{\phi}](x,x') = \delta(x-x') \,\{ L_S(x),L_I(x) \,\}$, with
\begin{subequations}\label{e:Ldef}
\begin{align}
   L_I(x)
   &=
   \bigl (\,
      \partial_t
      - 
      D_E \laplacian 
      +
      \mu \,
   \bigr ) \, \phi_E(x)
   -
   \lambda \, \phi_S(x) \, \phi_I(x) \,
   \label{e:LEdef} \\
   L_S(x)
   &=
   \bigl (\,
      \partial_t
      - 
      D_S \laplacian 
      +
      \nu \,
   \bigr ) \, \phi_S(x)
   +
   \lambda \, \phi_S(x) \, \phi_I(x)
   -
   f \,
   \label{e:LSdef} \\
\end{align}
\end{subequations}
and $\vb{D}[\vb*{\phi}](x,x') = \delta(x-x') \, \vb{D}[\vb*{\phi}](x)$ with
\begin{equation}\label{e:Ddef}
   \vb{D}[\vb*{\phi}](x)
   =
   \sigma(x) \,
   \begin{pmatrix}
      2 & -1 \\
      -1 & 0 
   \end{pmatrix}
   \qc
   \sigma(x) = - \lambda \, \phi_S(x) \, \phi_I(x) \>.
\end{equation}
Using the identity
\begin{align}\label{PI.e:18}
   &\sqrt{\det{\vb{D}}}
   \exp{
      \frac{1}{2}
      \iint \!\dd{x} \dd{x'}
      \vb*{\phi}^{\star}(x) \cdot \vb{D}[\vb*{\phi}](x,x') \cdot \vb*{\phi}^{\star}(x') }
   \\
   & \hspace{-1em}
   =
   \int \! \DD{\vb*{\eta}} \,
   \exp{ 
      - \frac{1}{2}
      \iint \!\dd{x} \dd{x'}
      \vb*{\eta}(x) \cdot \vb{D}^{-1}[\vb*{\phi}](x,x') \cdot \vb*{\eta}(x')
      +
      \int \dd{x} \, \vb*{\phi}^{\star}(x) \cdot \vb*{\eta}(x) 
         } \>,
   \notag
\end{align}
where we have set $\vb*{\eta}(x) = \{\, \eta_I(x),\eta_S(x) \, \}$,
the path integral becomes
\begin{equation}\label{PI.e:20}
   Z
   =
   \mathcal{N} \!\! \iiint \! 
   \DD{\vb*{\phi}^{\star}} \DD{\vb*{\phi}} \DD{\vb*{\eta}}
   \rme^{ - S[ \vb*{\phi},\vb*{\phi}^{\star},\vb*{\eta}  ] } \>,      
\end{equation}
where the action is now given by
\begin{align}\label{PI.e:21}
   \tilde{S}[\,\vb*{\phi},\vb*{\phi}^{\star},\vb*{\eta} \,]
   &=
   \iint\! \dd{x} \dd{x'}
   \Bigl \{\,
      \vb*{\phi}^{\star}(x) \cdot 
      [\, \vb{L}[\vb*{\phi}](x,x')
         -  
         \delta(x -x') \, \vb*{\eta}(x) \,
      ]
      \\
      & \hspace{2em}
      +
      \frac{1}{2} \,
      \vb*{\eta}(x) \cdot \vb{D}^{-1}[\vb*{\phi}](x,x') \cdot \vb*{\eta}(x) \,
   \Bigr \} \>.  
   \notag
\end{align}
Defining
\begin{equation}\label{PI.e:22}
   P[\vb*{\phi},\vb*{\eta}]
   =
   \mathcal{N} \,
   \exp{
      - \frac{1}{2} \,
      \vb*{\eta}(x) \cdot \vb{D}^{-1}[\vb*{\phi}](x,x') \cdot \vb*{\eta}(x) \,
        } \>,     
\end{equation}
the path integral \ef{PI.e:20} can be written as
\begin{align}\label{PI.e:23}
   Z
   &=
   \mathcal{N} \!\! \iint \! 
   \DD{\vb*{\phi}} \DD{\vb*{\eta}}
   P[\vb*{\phi},\vb*{\eta}]
   \\
   & \hspace{-1em}
   \times
   \int \! \DD{\vb*{\phi}^{\star}}
   \exp{
      - \iint\! \dd{x} \dd{x'}
      \vb*{\phi}^{\star}(x) \vdot 
      [\, \vb{L}[\vb*{\phi}](x,x')
         -
         \delta(x -x') \, \vb*{\eta}(x) \,
      ] }
   \notag 
\end{align}
The noise terms can be related to Gaussian noise by using a Cholesky decomposition.  We write the $\vb{D}[\vb*{\phi}](x)$ matrix as
\begin{equation}\label{PI.e:25}
   \vb{D}(x) = \vb{M}^{T}(x) \cdot \vb{M}(x) \>,
\end{equation}
where
\begin{equation}\label{PI.e:25}
   \vb{M}^{T}(x)
   =
   \sqrt{\frac{\sigma(x)}{2}} 
   \begin{pmatrix}
      2 & 0 \\
      - 1 & - \rmi 
   \end{pmatrix} \>.
\end{equation}
Writing
\begin{equation}\label{PI.e:26}
   \vb*{\eta}^{T}(x) \, \vb{D}^{-1}(x) \, \vb*{\eta}(x)
   =
   \vb*{\eta}^{T}(x) \, \vb{M}^{-1}(x) \, [\, \vb{M}^{T}(x) \,]^{-1} \, \vb*{\eta}(x)
   =
   \vb*{\theta}^{T}(x) \, \vb*{\theta}(x) \>,
\end{equation}
where we have put $\vb*{\theta}(x) = \{\, \theta_1(x),\theta_2(x), \theta_3(x) \,\}$. So $\vb*{\eta}(x) = \vb{M}^{T}(x) \, \vb*{\theta}(x)$, which gives
\begin{subequations}\label{e:noiseI}
\begin{align}
   \eta_I(x)
   &=
   \sqrt{2 \sigma(x)} \, \theta_1(x) \>,
   \label{e:noiseI-a} \\
   \eta_S(x)
   &=
   - \sqrt{\sigma(x)/2} \, 
   [\,\theta_1(x) + \rmi \, \theta_2(x) \,] \>.
   \label{e:noiseI-b}
\end{align}
\end{subequations}
The amplitude of the noise depends on the densities $\sigma(x) = - \phi_I(x) \phi_S(x)$.  
Noting that $\sqrt{\det{\vb{D}}} = \det{\vb{M}}$, the probability distribution function \ef{PI.e:22} becomes
\begin{equation}\label{PI.e:28}
   P[ \vb*{\theta} ]
   =
   P[\vb*{\phi},\vb*{\eta}] \, 
   \Big | \, \frac{\delta \vb*{\eta}}{\delta \vb*{\theta}} \, \Big |
   =
   \mathcal{N} \,
   \exp{ - \frac{1}{2} \int \!\dd{x} \vb*{\theta}^{T}(x) \, \vb*{\theta}(x) \,} \>,     
\end{equation}
which is uncorrelated white noise.  Changing integration variables from $\vb*{\eta}$ to $\vb*{\theta}$, the path integral \ef{PI.e:23} becomes:
\begin{equation}\label{PI.e:29}
   K
   =
   \mathcal{N} \!\! \iint \! 
   \DD{\vb*{\phi}} \DD{\vb*{\theta}}
   P[\vb*{\theta}] \,
   \big | \det{\vb{M}[\vb*{\phi}]} \big | \,
   \delta[\, \vb{L}[\vb*{\phi}](x) - \vb*{\eta}(x) \,] \>,
\end{equation}
which has value along a path defined by $\vb{L}[\vb*{\phi}](x) = \vb*{\eta}(x)$, given by the Langevin  equations,
\begin{subequations}\label{e:Langevin}
\begin{align}
   \bigl (\,
      \partial_t
      - 
      D_E \laplacian 
      +
      \mu \,
   \bigr ) \, \phi_E(x)
   -
   \lambda \, \phi_S(x) \, \phi_I(x)
   &=
   \eta_I(x) \>,
   \label{e:Langevin-a} \\
   \bigl (\,
      \partial_t
      - 
      D_S \laplacian 
      +
      \nu \,
   \bigr ) \, \phi_S(x)
   +
   \lambda \, \phi_S(x) \, \phi_I(x)
   -
   f
   &=
   \eta_S(x) \>,
   \label{e:Langevin-b}
\end{align}
\end{subequations}
with noise functions given by \ef{e:noiseI}.

%
%
\section{\label{s:random}Gaussian random variables}

The $\theta_i(x)$ in Eqs.~\ef{e:noiseI} are random variables deriverd from independent Gaussian white noise distributions with zero mean and unit variance.  They are defined by the path integral:
\begin{equation}\label{GN.e:1}
   P[\, \vb*{\theta} \,]
   =
   \mathcal{N} \exp( 
      - \frac{1}{2} \int \dd[d]{x} \! \int \dd{t} 
      \vb*{\theta}^T(\vb{x},t) \vdot \vb*{\theta}(\vb{x},t) ) \>.
\end{equation}
Expanding the noise function on a space-time lattice defined by $\vb{x}_{\vb{i}} = h \, \vb{i}$ and $t = \Delta t \, j$, Eq.~\ef{GN.e:1} becomes
\begin{align}\label{GN.e:2}
   P[\, \vb*{\theta} \,]
   &=
   \mathcal{N} \exp( 
      - \frac{h^d \Delta t}{2} 
      \sum_{\vb{i},j} \theta_{\vb{i},j}^2 )
   \\
   &=
   \mathcal{N} \prod_{\vb{i},j} \,
   \exp( - \frac{h^d \Delta t}{2} \, \theta_{\vb{i},j}^2 ) \>.
   \notag
\end{align}
So now setting
\begin{equation}\label{GN.e:3}
   \tilde{\theta}_{\vb{i},j}
   =
   \sqrt{h^d \Delta t} \, \theta_{\vb{i},j} \>,
\end{equation}
then $\tilde{\theta}_{\vb{i},j}$ is \emph{dimensionless} and taken from a normal distribution with zero mean and unit variance:
\begin{equation}\label{GN.e:4}
   P[\, \tilde{\theta} \,]
   =
   \frac{1}{\sqrt{2 \pi}} \, \exp( - \frac{\tilde{\theta}^2}{2} ) \>,
\end{equation}
for all $\tilde{\theta}_{\vb{i},j}$.  So for a $\Delta t$ step,
\begin{equation}\label{GN.e:5}
   \dd{\theta_{\vb{i},j}}
   \equiv
   \theta_{\vb{i},j} \Delta t
   =
   \sqrt{\frac{\Delta t}{h^d}} \, \tilde{\theta}_{\vb{i},j} \>.
\end{equation}
That is, the change in noise currents is proportional to $\sqrt{\Delta \tilde{t}}$, which is a Weiner process.  This is a result of the statistics.

%
%
\section{\label{s:micro}Microscopic equations}

There are different ways to implement a stochastic method for directly simulating the master equation.  One method is to choose a very small but fixed time step and using a probability function and random numbers to decide for each step which reaction takes place, starting from some initial condition.  The process is repeated starting from the same initial conditions and ensemble averages computed.  In 1976, Gillispie \cite{GILLESPIE1976403,Gillespie:1977aa} realized that the problem with this method was that for most of the steps, nothing happens.  So if one could estimate the value of a time step in which nothing happens, one could take steps when something really happens.  In addition, updates of occupation numbers at all sites can be done in single pass.  The way this works is the following.  Let $a_i(\vb{n})$ be the reaction rates for the $i^{\text{th}}$ process including a jump process, so that $a(\vb{n}) = \sum_{i=1}^{q} a_i(\vb{n})$ is the reaction probability per unit time that something happens.  So then the probability that nothing happens in a time between $t$ and $t + \Delta t$ is given by
\begin{equation}\label{ptime}
   P(\Delta t) = 1 - a(\vb{n}) \, \Delta t + \dots = \rme^{- a(\vb{n}) \, \Delta t} \>,
\end{equation}
where $P(\Delta t) = r$, where $r$ is a random number in the range $0 \le r \le 1$.  Solving for $\Delta t$ gives the Gillispie algrothm:
\begin{equation}\label{Gillispie}
   \Delta t = \frac{ \ln(1/r) }{a(\vb{n})} \>.
\end{equation}
Selecting a random number $r$ then gives the value of the time step.  
One is guaranteed (more or less) that during this time, no interactions will take place.  So one can then take this step and then decide, using new random numbers, which process \emph{do} take place.  The algorithm then update the occupation numbers using the stochastic vectors of the reaction for all the reaction elements at all sites.  Then one recomputes $a(\vb{n})$ and find a new step to take.  Specifically, the Gillespie algorithm is the following steps \cite{erban2007practical}:
\begin{enumerate}
\item Set the time $t=0$ and the initial occupation values for the reactants $n_1(0)$.
\item Generate two random numbers $r_1$ and $r_2$ uniformly distributed in $(0,1)$.
\item Compute the propensity function $a_i(\vb{n})$ for each reaction and the sum over all reactions: 
\begin{equation}\label{asum}
   a(\vb{n}) = \sum_{i=1}^{q} a_i(\vb{n}) \>.
\end{equation}
\item Using \ef{Gillispie}, compute the time $\Delta t$ when the next reaction takes place and update the time, $t = t + \Delta t$.  
\item Compute \emph{which} reaction takes place by finding the value of $j$ where
\begin{equation}\label{jvalue}
   \frac{1}{a(\vb{n})} \sum_{i=1}^{j-1} a_i(\vb{n}) \le r_2 < \frac{1}{a(\vb{n})} \sum_{i=1}^{j} a_i(\vb{n}) \>.
\end{equation}
Then the $j^{\text{th}}$ reaction takes place.  Since $r_2$ is a random number it doesn't matter how the reactions are ordered.  Then update the occupation numbers of reactants and products of the $j^{\text{th}}$ reaction.
\item Go back to step 2 and continue to the end time.
\end{enumerate}
There is only one reaction per time step so one does not have to recompute all the propensity functions at each time step, but can just update only those that are changed by the reaction selected.  
Reaction parameters for the SEIR model are given in Table~\ref{t:Reactions}.

Diffusion can be included by creating a space region of $0 \le x \le L$ and a grid of $\Delta x = L/N$.  Reactants are now labeled by the space region $i$ and the reactant $\alpha$.  Reactions take place only at the same location, but can jump to neighboring locations at a rate $d_{\alpha}$.  The propensity function for jumping is $a_{i,\alpha} = d_{\alpha} n_{i,\alpha}$.  In one-dimension, jumps can occur to the right or the left at (usually) equal rates.  Stochastic vectors are $s_{\alpha} = \pm 1$.  Reflective boundary conditions are used.  We are in the process of implementing this method using a modification of a Fortran code by Erban, Chapman, and Maini \cite{erban2007practical,erban_chapman_2020} (See also \url{http://www.maths.ox.ac.uk/cmb/Education/}) which uses the Gillespie algorithm.

%
%
\bibliography{/Users/dawson/Dropbox/Utilities/johns.bib}
%
%
\end{document}